\documentclass[%
 aip,
 amsmath,amssymb,
 reprint,%
]{revtex4-1}

\usepackage{graphicx}% Include figure files
\usepackage{dcolumn}% Align table columns on decimal point
\usepackage{mathptmx}
\usepackage[utf8]{inputenc} % allow utf-8 input
\usepackage[T1]{fontenc}    % use 8-bit T1 fonts
\usepackage{hyperref}       % hyperlinks
\usepackage{url}            % simple URL typesetting
\usepackage{booktabs}       % professional-quality tables
\usepackage{amsfonts}       % blackboard math symbols
\usepackage{nicefrac}       % compact symbols for 1/2, etc.
\usepackage{microtype}      % microtypography
\usepackage{lipsum}
\usepackage{graphicx}
\usepackage{subfigure}
\usepackage{multirow}
\usepackage{makecell}
\usepackage{url}
\usepackage[toc, page]{appendix}
\usepackage{bm}

\begin{document}

\preprint{AIP/123-QED}

\title{Robust active flow control over a range of Reynolds numbers using an
artificial neural network trained through deep reinforcement learning}
% Force line breaks with \\

\author{Hongwei Tang}
 \affiliation{College of Aerospace Engineering, Nanjing University of Aeronautics and Astronautics\\
 29 Yudao St., Nanjing 210016, China}
 \altaffiliation[Also at ]{Jiangsu Key Laboratory of Hi-Tech Research for Wind Turbine Design}

\author{Jean Rabault}%
\affiliation{Department of Mathematics, University of Oslo, 0316 Oslo, Norway}%
\altaffiliation[Also at ]{MINES Paristech , PSL Research University, CEMEF}

\author{Alexander Kuhnle}
\affiliation{Blue Prism AI Labs, London WC2B 6NH, United Kingdom}%

\author{Yan Wang}
\affiliation{College of Aerospace Engineering, Nanjing University of Aeronautics and Astronautics\\
29 Yudao St., Nanjing 210016, China}
 \altaffiliation[Also at ]{Jiangsu Key Laboratory of Hi-Tech Research for Wind Turbine Design}

 \author{Tongguang Wang}
 \email{tgwang@nuaa.edu.cn}
 \affiliation{College of Aerospace Engineering, Nanjing University of Aeronautics and Astronautics\\
 29 Yudao St., Nanjing 210016, China}
 \altaffiliation[Also at ]{Jiangsu Key Laboratory of Hi-Tech Research for Wind Turbine Design}

\date{\today}% It is always \today, today,

\begin{abstract}
  This paper focuses on the active flow control of a computational fluid
  dynamics simulation over a range of Reynolds numbers using deep reinforcement
  learning (DRL). More precisely, the proximal policy optimization (PPO) method
  is used to control the mass flow rate of four synthetic jets symmetrically
  located on the upper and lower sides of a cylinder immersed in a
  two-dimensional flow domain. The learning environment supports four flow
  configurations with Reynolds numbers 100, 200, 300 and 400, respectively. A
  new smoothing interpolation function is proposed to help the PPO algorithm to
  learn to set continuous actions, which is of great importance to effectively
  suppress problematic jumps in lift and allow a better convergence for the
  training process. It is shown that the DRL controller is able to significantly
  reduce the lift and drag fluctuations and to actively reduce the drag by
  approximately 5.7\%, 21.6\%, 32.7\%, and 38.7\%, at $Re$=100, 200, 300, and
  400 respectively. More importantly, it can also effectively reduce drag for
  any previously unseen value of the Reynolds number between 60 and 400. This
  highlights the generalization ability of deep neural networks and is an
  important milestone towards the development of practical applications of DRL
  to active flow control.
\end{abstract}

\maketitle

\section{Introduction}

Actively controlling a flow to change its characteristics is attractive for many
applications in the field of fluid mechanics and could bring large industrial
benefits \cite{gad2000flow}. Since the pioneering work of Prandtl about the use
of active flow control (AFC) for delaying boundary layer separation
\cite{prandtl1904}, AFC has witnessed a fast growth and has become an
increasingly important technology for the pursuit of industrial and sustainable
solutions \cite{articleRVinu}. Prospective applications of AFC to problems of
industrial and environmental importance include, to name a few, reducing the
aerodynamic drag on aircrafts \cite{Shahrabi2019, doi:10.2514/1.J056258},
manipulating the vortex in the wake of bluff bodies \cite{Zhu2019, Wang2017,
doi:10.1063/1.5109320, WANG2016160, doi:10.1063/1.5127202} and optimizing the
design and performance of wind turbines \cite{Aubrun2017, doi:10.1002/we.2109,
doi:10.1002/we.1737} and gas turbines \cite{doi:10.2514/1.J056697}. 

Nevertheless, finding efficient strategies for performing AFC remains a
challenge \cite{gad2000flow,Brunton2020}. This difficulty is deeply rooted in
the nature of the Navier-Stokes equations and their underlying high
non-linearity, as well as in the high dimensionality of possible control
parameter spaces. Additionally, considerable challenges exist for applying AFC
to engineering situations, such as disturbances inherent to the physical
environment, and imperfections in the manufacturing or installing of the
actuators, which impose hard requirements on the ability of control algorithms
to adapt robustly to external conditions. This makes the design of control
strategies a complex endeavour. Therefore, the main issue of AFC is currently
the lack of robust, efficient algorithms that can leverage the physical devices
available for performing effective control.

In practice, AFC can be open-loop (no feedback mechanism) or closed-loop (when a
feedback mechanism is present, i.e. some measurements of the flow are provided
to the AFC system to decide the next actuation) \cite{collis2004issues}.
Compared with open-loop control, closed-loop control possesses more potential to
take full advantage of active devices to alter the flow. At present, many
implementations of AFC are based on mathematical models of the flow system. For
example, Flinois \textit{et al.} \cite{doi:10.1063/1.4928896} developed an
adjoint-based optimal control framework to help stabilize the vortex shedding
efficiently. Leclercq \textit{et al.} \cite{leclercq_2019} proposed a
feedback-loop strategy using iteratively linearized models to suppress
oscillations of resonating flows. Bergmann \textit{et al.} \cite{Bergmann2005}
deduced an optimal control approach for the flow past a circular cylinder using
proper orthogonal decomposition reduced-order models. Brackston \textit{et al.}
\cite{cruz_wynn_rigas_morrison_2016} used a stochastic modelling approach to
design a feedback controller and validated it in experiments, effectively
suppressing the asymmetric large-scale structure behind a bluff body wake with active
flaps. These model-based control strategies are usually based on either harmonic
or constant forcing \cite{brunton2015closed, doi:10.1063/1.869789}, making it
however challenging for real-world AFC where complex non-linear systems are
present in combination with stochastic disturbances
\cite{gautier_aider_duriez_noack_segond_abel_2015}.

By contrast, model-free approaches, where the control strategy is found through
a data-driven and learning-based approach, are quite suitable for complex,
high-dimensional, nonlinear systems \cite{Brunton2020,brunton2015closed}. Such
techniques mainly include genetic algorithms (GAs) and artificial neural
networks (ANNs). While GAs have been extensively used for AFC
\cite{duriez2017machine, doi:10.1063/1.5115258,
gautier_aider_duriez_noack_segond_abel_2015, debien2016closed}, ANNs are
receiving growing attention recently due to the fast development of artificial
intelligence / machine learning that has taken place in recent years.
Furthermore, ANNs have been found so far to surpass GAs in terms of the
complexity of the tasks learned and their learning speed \cite{Mnih2015,
duan2016benchmarking}. Among other methods within the field of machine learning,
ANNs used together with reinforcement learning algorithms have attracted great
attention \cite{gu2016continuous, hessel2018rainbow}. The resulting deep
reinforcement learning (DRL) paradigm has been successfully deployed to resolve
several high-profile, complex problems, like playing a wide range of Atari game
without hard-coding strategies \cite{mnih2013playing}, generating realistic
dialogues \cite{li2016deep}, or controlling the dynamics of complex robots
\cite{gu2017deep}. Compared with data-driven and supervised learning approaches,
which have also found some applications in fluid mechanics within particle image
velocimetry (PIV) measurement \cite{Rabault_2017, MENDEZ2018256, MENDEZ201948}, reduced-order modeling
\cite{mendez_balabane_buchlin_2019,
doi:10.1063/1.5144861}, or predictions of flow features
\cite{guastoni2019prediction, doi:10.1063/1.5094943, doi:10.1063/1.5109698}, DRL
allows to find a solution through trial-and-error, even when no solution is
known a-priori. One can observe that challenging systems successfully controlled
by DRL have remarkably similar properties of nonlinearity and high-dimension,
similar to the features of flow phenomena that make AFC challenging.
Consequently, DRL is seen as a promising avenue for performing AFC
\cite{Brunton2020}.

Therefore, in recent years, DRL has became a new tool to discover AFC strategies
\cite{Brunton2020}, and it has been shown to outperform previous
techniques in several cases \cite{PhysRevFluids.4.093902}. In addition,
increases in the computational power available for numerical simulations make it possible
to study increasingly complex systems using DRL and simulations. Such applications include
optimizing the motion for individual \cite{doi:10.1137/130943078} or collective
fishes \cite{brauer_koumoutsakos_2016, Verma5849}, training a glider to
autonomously navigate atmospheric thermic current \cite{reddy2018glider}, and
controlling the adaptive behavior of microswimmers
\cite{PhysRevLett.118.158004}. Although the computational costs of the
simulations needed to train the DRL algorithms still limit their application,
they already have helped shed light on several complex problems.

The present work is an extension of the results initially presented by Rabault
\textit{et al.} \cite{Rabault2019a, Rabault2019b}, but with four synthetic jets
which are located symmetrically on a cylinder immersed in a two-dimensional
domain. Moreover, the ability of DRL to
design robust active control strategies for the flow over a range of conditions is further investigated.
The PPO agent together with a 2-layer fully connected neural network is used
to control the mass flow rates of these four jets to reduce the magnitude and
oscillation of the drag. In addition, a new interpolation equation is developed to make the control
values change smoothly in time so that problematical lift oscillations, which
are caused by the interpolation function proposed in previous works \cite{Rabault2019a,
Rabault2019b}, are almost completely eliminated. In addition, the robustness
and feasibility of the obtained control strategy which shows the best
performance in different flow conditions is discussed. The paper is organized as follows.
Firstly, a brief introduction to the
numerical method used for performing the simulations, and the general theory
underlying the DRL algorithm used, is provided in Sec.~\ref{sec:methods} . The
training using the DRL algorithm over a
range of Reynolds numbers is then detailed in Sec.~\ref{sec:results}, together
with the results which underline the robustness and generalization ability of
the control strategy obtained. Finally, a brief summary of the contribution and
its significance for the use of DRL within AFC are demonstrated in Sec.~\ref{sec:conclusions}.

\section{Problem set-up and methodology}\label{sec:methods}

\subsection{Problem description}\label{sec:geometry}

\begin{figure*}
  \centering
  \includegraphics[width=13cm, height=4cm]{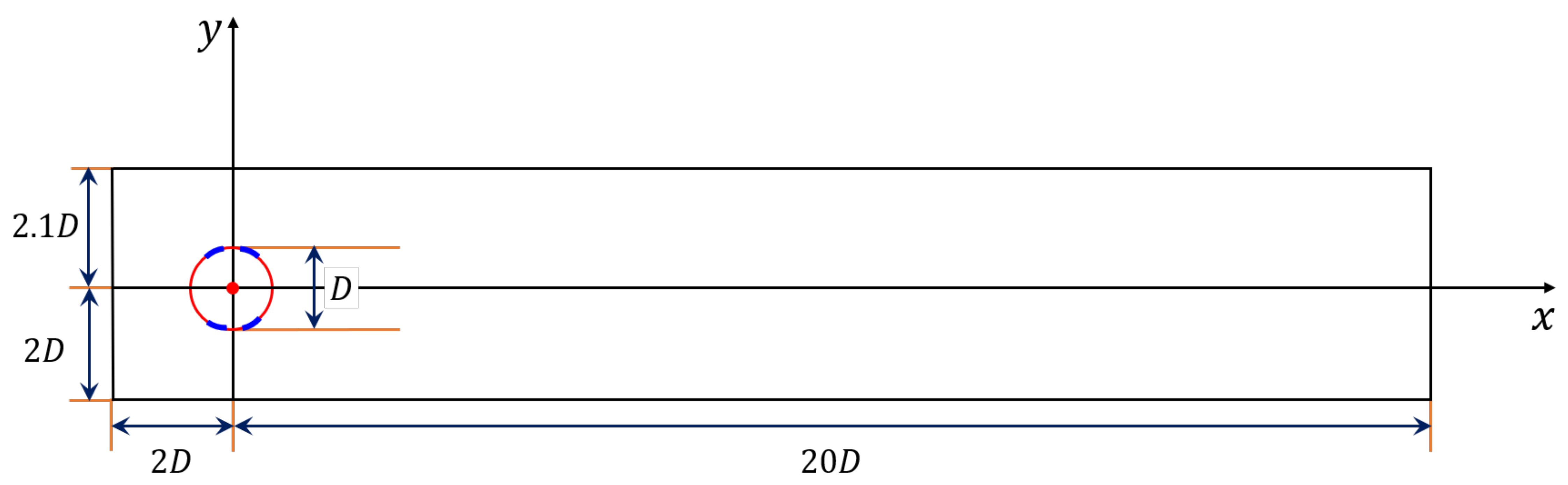}
  \caption{Geometrical description of configuration used for simulating the flow
    past a circular cylinder immersed in a two-dimensional channel, adapted from
    the work of Sch\"afer \textit{et al.} \cite{Schafer1996}. The center of the
    cylinder and the synthetic jets are marked by red dot and blue arcs,
    respectively. The cylinder is slightly off the horizontal centerline of the
    channel (by $0.05D$). This geometric asymmetry helps trigger the
    vortex shedding.}
  \label{flow_geometry}
\end{figure*}

The configuration of the simulation is adapted from the classical benchmark
computations carried out by Sch\"afer \textit{et al.} \cite{Schafer1996} (also
known as the Turek benchmarks), in which a cylinder of diameter $D$ is immersed
in a two-dimensional domain with size $22D\times 4.1D$, as depicted in Fig.
\ref{flow_geometry}. The center of the cylinder is located at a transversal
distance of $0.05D$ from the horizontal centerline of the flow domain.
This geometric asymmetry helps trigger the vortex shedding if the
Reynolds number is greater than the critical value.

For performing AFC, four jets, for which the mass flow rates are controlled by
the ANN, are symmetrically located on the upper and lower sides of the cylinder.
The angular positions of these four jets are $75^\circ$ (corresponding to
$\theta_0$ as shown in Fig.~\ref{BC}), $105^\circ$, $255^\circ$ and
$285^\circ$, respectively. The jets are chosen as synthetic jets, i.e., the sum
of the mass flow rates of all jets is enforced to be zero, and the jet
directions are set to be perpendicular to the cylinder wall. The injection
velocity can be positive or negative, corresponding to blowing or suction,
respectively. With such configurations, there could be extra injected momentum
that could act as propulsion, as discussed in Appendix \ref{appendix_b}.
However, the propulsion is in any case small thanks to the net
mass flow rate being kept equal to zero, and it amounts for no more than 5\% of the momentum
intercepting the cylinder once a pseudo-periodic regime with active control has
been achieved. Therefore, this small propulsion effect will be neglected in the
following discussion.

\subsection{Numerical method}\label{sec:solver}

In the present study, the flow is assumed to be viscous and incompressible. The
governing equations are the two-dimensional, time-dependent Navier-Stokes
equations and the continuity equation, which can be expressed in non-dimensional
form as:

\begin{equation}
  \frac{\partial\bm u}{\partial t} + \bm u\cdot(\nabla\bm u) = -\nabla p + \frac{1}{Re}\Delta\bm u,
  \label{NS}
\end{equation}
\begin{equation}
  \nabla\cdot\bm u=0,
  \label{continuity}
\end{equation}

\noindent where $\bm u$ is the non-dimensional velocity, $t$ is the
non-dimensional time, $p$ is the non-dimensional pressure. The characteristic
length, velocity, density and time for non-dimensionalizing the problem are $D$,
$\overline U$, $\rho$, and $D/\overline U$, respectively, where $\overline U$ is
the bulk velocity as will be shown later, and $\rho$ is the density of the
fluid. The Reynolds number is defined as $Re=\overline{U}D/\nu$, where $\nu$ is
the kinematic viscosity of fluid.

\begin{figure}
  \centering
  \includegraphics[width=8cm, height=5.5cm]{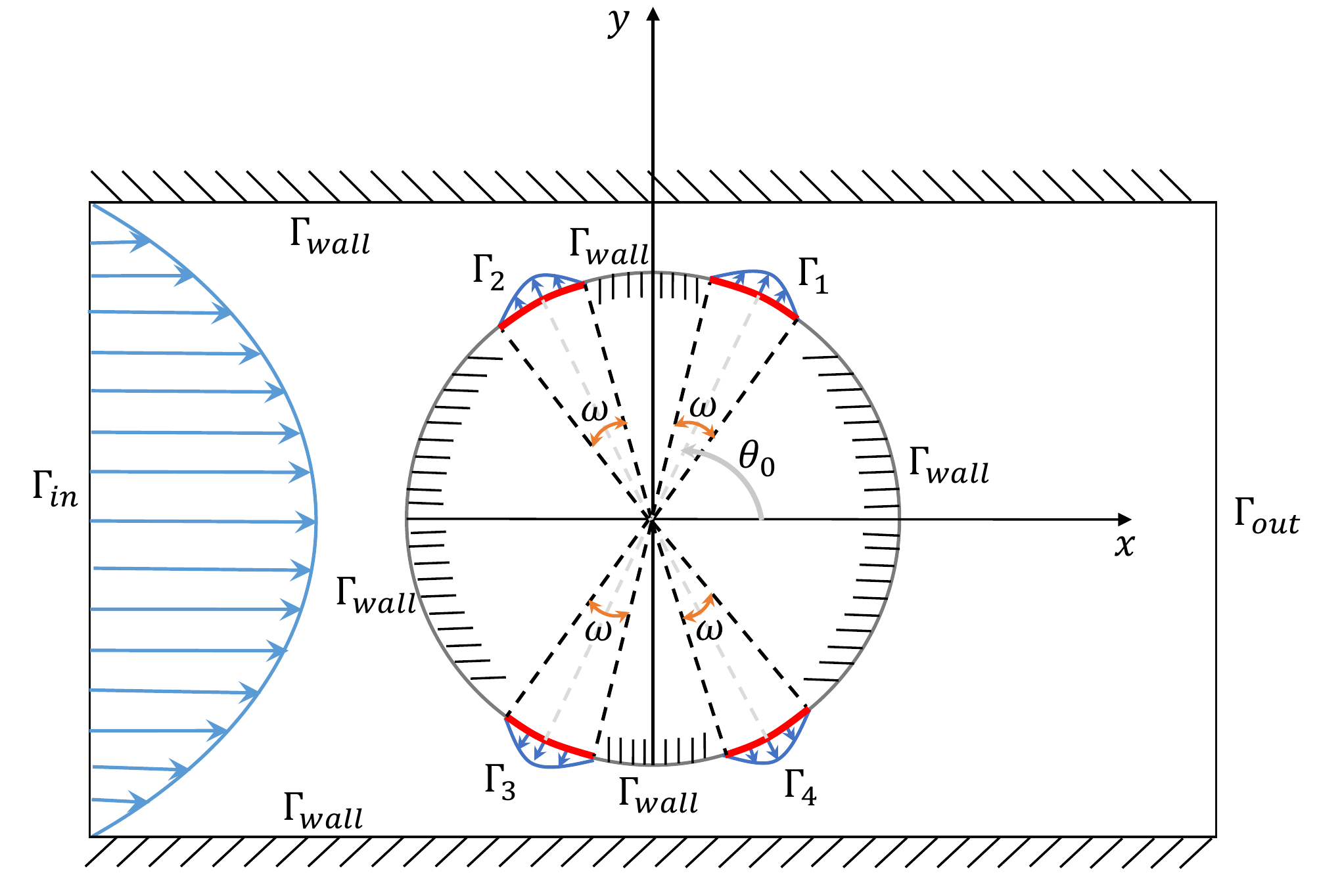}
  \caption{Flow domain (not at scale) and boundary conditions for the
  simulation. The jet velocity profiles, determined by the output of ANNs, are
  prescribed (red arcs) by $\Gamma_j$ ($j=1, 2, 3, 4$).
  $\Gamma_{wall}$ means no-slip boundary conditions implemented for solid walls.
  $\Gamma_{in}$ is the inflow part while $\Gamma_{out}$ represents outflow.
  $\omega$ is the width of the jets.}
  \label{BC}
\end{figure}

Fig.~\ref{BC} shows a schematic of the boundary conditions (for illustration
purpose, the geometrical domain is out of scale). The inflow velocity profile in
the streamwise direction ($\Gamma_i$) is specified as (cf. 2D-2 test case
reported by Sch\"afer \textit{et al.} \cite{Schafer1996}):

\begin{equation}
  u_{inlet}(y)=-4U_m(y-2.1D)(y+2D)/H^2,
\end{equation}

\noindent where $H=4.1D$ is the width (along the $Y$-axis as depicted in Fig.
\ref{flow_geometry}) of the domain, and $U_m$ is the horizontal velocity component at
the midpoint of the inlet, i.e., the maximum of the inflow velocity. As a
consequence, the bulk velocity can be calculated as:

\begin{equation}
  \overline U=\frac{1}{H}\int_{-2D}^{2.1D}u_{inlet}(y)dy=\frac{2}{3}U_m.
\end{equation}

No-slip boundary conditions ($\Gamma_{wall}$), i.e., the velocity of fluid is
zero, are applied on the top and bottom walls and on the solid walls of the
cylinder. The boundary condition corresponding to an out-flow boundary
($\Gamma_{out}$) is imposed based on the assumption that the derivative of the
velocity along the $X$-axis is zero at the outlet, which implies that the flow
is fully-developed or does not change significantly. More strictly, it is set
as:

\begin{equation}
  -p\bm n+\frac{1}{Re}(\nabla\bm u\cdot\bm n)=0,
\end{equation}

\noindent where $\bm n$ is the unit vector normal to the outlet.

To avoid velocity discontinuity between the boundary of the jets and the no-slip
surfaces of the cylinder, the radial velocity profiles ($\Gamma_j$) of the four
synthetic jets are prescribed as:

\begin{equation}
  u_{jet}(\theta, Q_i)=\frac{\pi}{\omega D}Q_i\cos(\frac{\pi}{\omega}(\theta-\theta_0)),
\end{equation}

\noindent where $Q_i(i=1, 2, 3, 4)$ is the mass flow rate of the four jets centered at
$\theta_0=75^\circ$, $105^\circ$, $255^\circ$ and $285^\circ$, respectively.
$\omega=10^\circ$ is the width of each jet.

For solving Eqs.~\ref{NS} and \ref{continuity} numerically, the incremental
pressure correction scheme (IPCS) method \cite{Goda1979} is used with explicitly
linearization of the nonlinear convective term by using the known velocity $\bm
u^n$ at time step $t=n\delta t$, where $\delta t$ is the numerical timestep and
$n$ is the number of the timestep considered. This method is applied as a
two-step fractional step method. First, an auxiliary velocity
$\hat{\bm u}$ is calculated by:

\begin{equation}
  \frac{1}{\delta t}(\hat{\bm u}-\bm u^n)=-\bm u^n\cdot(\nabla\bm u^n)-\nabla p^n+\frac{1}{Re}\Delta\frac{\hat{\bm u}+\bm u^n}{2},
\end{equation}

\noindent then the pressure $p^{n+1}$ at $t=(n+1)\delta t$ is obtained by
solving a Poisson equation:

\begin{equation}
  \Delta (p^{n+1}-p^n)=\frac{1}{\delta t}\nabla\cdot\hat{\bm u}.
\end{equation}

This second step is usually referred to as the projection step.

Finally, the velocity $\bm u^{n+1}$ at $t=(n+1)\delta t$ is obtained by:

\begin{equation}
  \frac{1}{\delta t}(\bm u^{n+1}-\hat{\bm u})=-\nabla(p^{n+1}-p^n).
\end{equation}

\begin{figure*}
  \centering
  \subfigure[]{
  \includegraphics[width=12.5cm, height=2.8cm]{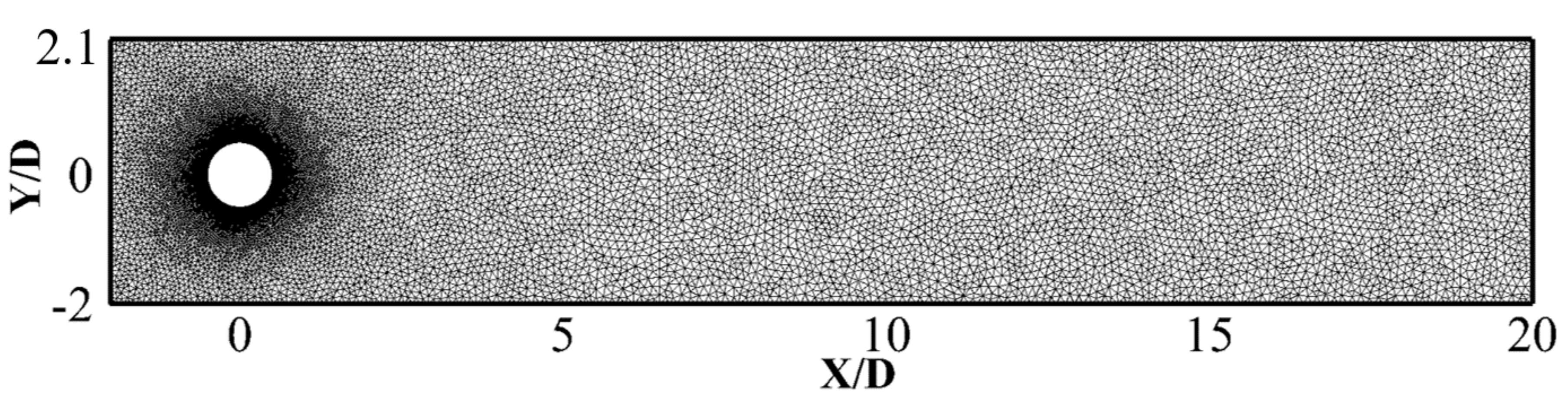}
  }
  \subfigure[]{
  \includegraphics[width=12cm, height=4.5cm]{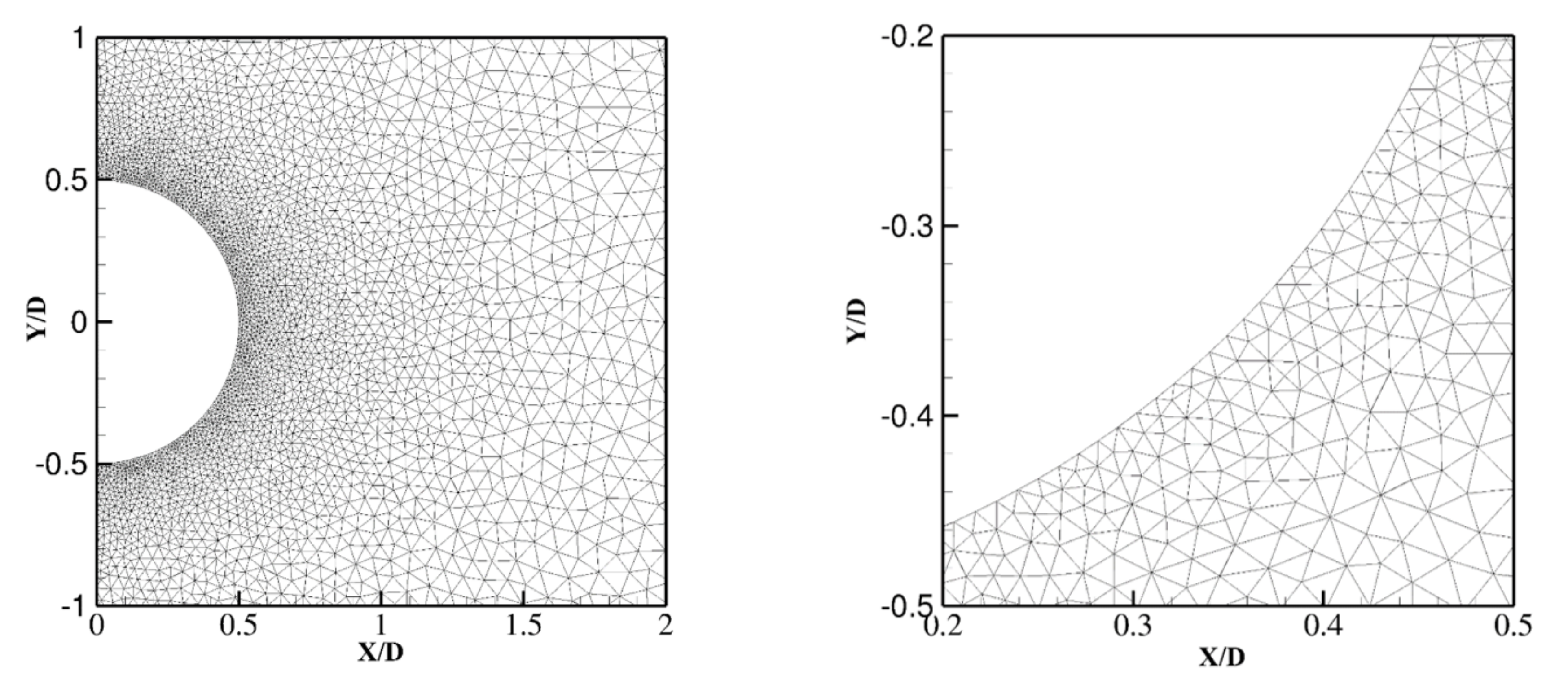}
  } \caption{Numerical discretization of the full (a) and partial (b)
  computational domain. The mesh is much refined around the cylinder to fully
  consider the influence of actuations on the flow simulation.}
\label{mesh}
\end{figure*}

The computational domain is discretized by an unstructured mesh (triangular
cells) and it is much refined around the surface of the cylinder (as shown in
Fig.~\ref{mesh}) so that the influence of synthetic jets on the flow simulation
can be fully considered. The IPCS method is implemented using the finite element
method within the FEniCS framework \cite{Logg2012}. More precisely, the linear
and quadratic basis functions of the continuous Galerkin family of elements are
utilized to discretize the pressure and velocity fields, respectively. The
resulting system of equations are solved using LU decomposition, a sparse direct
solver from the UMFPACK library \cite{Davis:1997:UMM:258211.258225}. The
numerical solution is obtained at each time step, and then the drag $F_D$ and
lift $F_L$ are integrated over the whole wall (including the jet surfaces) of
the cylinder by:

\begin{equation}
  F_D=\int(\bm\sigma\cdot\bm n_c)\cdot\bm e_xdS,
\end{equation}

\noindent and

\begin{equation}
  F_L=\int(\bm\sigma\cdot\bm n_c)\cdot\bm e_ydS,
\end{equation}

\noindent where $\bm\sigma$ is the Cauchy stress tensor, $\bm n_c$ is the unit
vector normal to the outer cylinder surface, and $\bm e_x=(1, 0)$, $\bm e_y=(0,
1)$.

In order to study the mesh convergence and validate the numerical method, the
quantities of interest are calculated from simulating the flow at $Re=100$, and
compared with the benchmark data \cite{Schafer1996}. The drag $F_D$ and
lift $F_L$ are normalized following:

\begin{equation}
  C_D=\frac{2F_D}{\rho \overline{U}^2D},
\end{equation}

\noindent and

\begin{equation}
  C_L=\frac{2F_L}{\rho \overline{U}^2D}.
\end{equation}

The Strouhal number ($St$), which is used to describe the characteristic
frequency of oscillating flow phenomena, is defined as:

\begin{equation}
  St=f_s\cdot D/\overline{U},
\end{equation}

\noindent where $f_s$ is the shedding frequency computed from the periodic evolution
of lift coefficient $C_L$.

The simulation results using meshes of three different resolutions are listed in
Tab.\ \ref{validation}, together with comparison to the bounds suggested by
Sch\"afer \textit{et al.} \cite{Schafer1996}. The $C_D^{max}$ and $C_L^{max}$
correspond to the maximum of the drag coefficient $C_D$ and lift coefficient $C_L$,
respectively. As can be seen, the resolution of main mesh, which is used in the
present work, is fine enough for the simulation to agree well with the benchmark
data. The discrepancies are less than 0.04\% in all listed quantities when
compared with fine mesh. Although the maximum of $C_L$ with the main mesh is
slightly larger than the suggested upper bound by approximately 2.2\%, the
discrepancy is small. Moreover, the maximum of $C_D$ and $St$ are strictly
within the suggested interval, which is of great importance as reducing drag is
the main focus. Hence, the main mesh depicted in Fig.\ \ref{mesh}
is deemed sufficiently refined and is used thereafter.

\begin{table*}[]
  \caption{Mesh convergence and flow parameters for the 2D flow around a circular cylinder at $Re=100$,
  in a configuration corresponding to the benchmark \cite{Schafer1996}.}
    \centering
  \label{validation}
  \begin{tabular}{@{}llrccc@{}}
  \toprule
  Case   &  & Mesh resolution & $C_D^{max}$ & $C_L^{max}$ & $St$ \\ \midrule
  \multirow{3}{*}{Present} & Coarse & 9374 & 3.2416 & 1.0758 & 0.3025 \\
                    & Main & 25865 & 3.2299 & 1.0323 & 0.3020 \\
                    & Fine & 174520 & 3.2311 & 1.0324 & 0.3020 \\
  Sch\"afer \textit{et al.}\cite{Schafer1996} &  & & 3.2200$\sim$3.2400 & 0.9900$\sim$1.0100 & 0.2950$\sim$0.3050 \\
  \bottomrule
  \end{tabular}%
  \end{table*}

\subsection{DRL control algorithm}\label{sec:DRL}

Advances in machine learning have promised a renaissance in understanding
intrinsic features of many complex systems and gain unprecedented attention not
only in computer science but also in many other disciplines, such as fluid
mechanics \cite{Kutz2017, Thuerey2019, Beck2019}, partial differential equations
\cite{Sirignano2018, Raissi2019}, or design optimization \cite{Yan2019,
Yonekura2019}. Reinforcement learning is one of the main branches of machine
learning and recently attracted a lot of interest following Google DeepMind
defeating top human professionals at the game of Go \cite{Silver2016}. Unlike
other machine learning methods such as supervised learning which consists in
learning to map an input to its corresponding output based on labeled examples provided
by a knowledgeable external supervisor, or unsupervised learning which is
typically interested in finding transformations and clustering properties hidden
in data, reinforcement learning is concerned with how to interact with an
environment so as to maximize a numerical reward signal.

\begin{figure*}
  \centering
  \includegraphics[width=16cm, height=6cm]{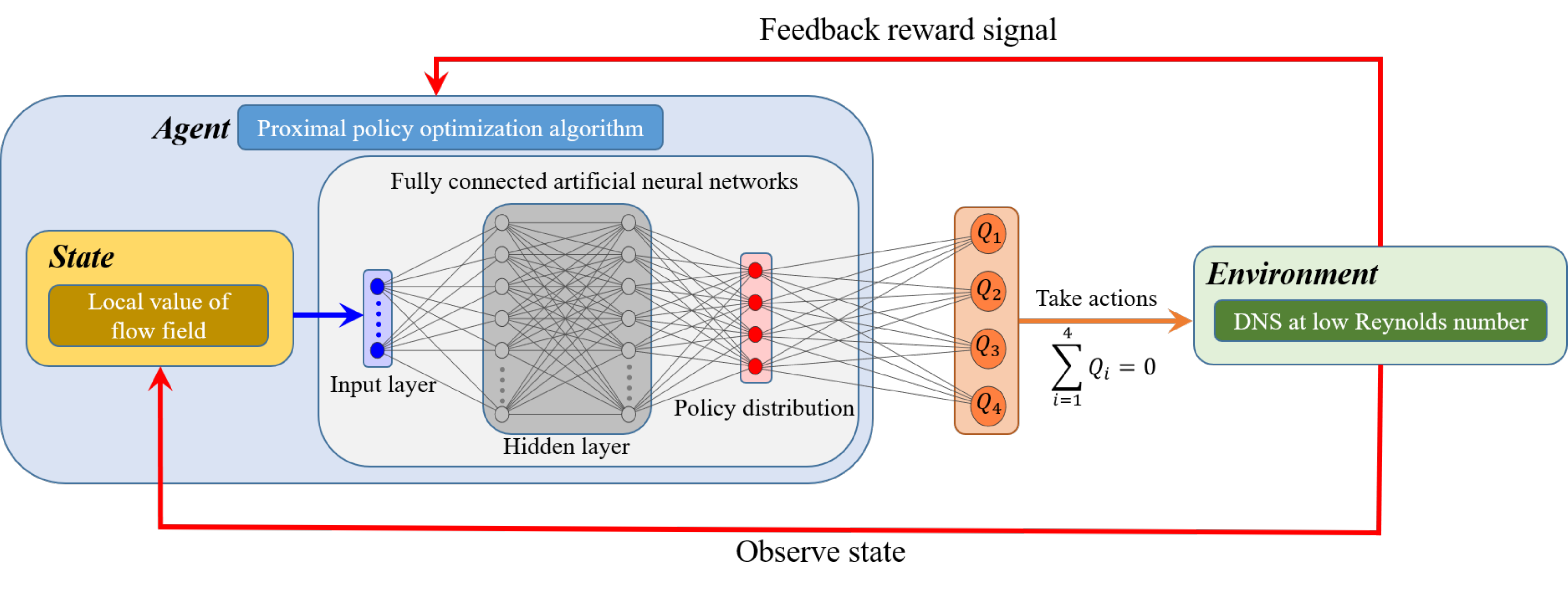}
  \caption{Illustration of the DRL framework utilized in the present work for
    performing AFC. The environment, i.e., a numerical simulation of the flow
    past a cylinder, is coupled in a closed-loop fashion with the learning
    agent. Iteratively, the mass flow rate of the jets ($Q_i(i=1, 2, 3, 4)$) is
    controlled by the agent according to the observed flow state. In response,
    the simulation produces the updated flow field as next state, and a
    reward signal is used to guide the control strategy towards controlling the
    flow so as to reduce the drag. Via such coupled interaction the agent
    eventually learns to perform effective AFC of the simulated flow.}
  \label{RL_framework}
\end{figure*}

\begin{figure*}
  \centering
  \includegraphics[width=13cm, height=2.5cm]{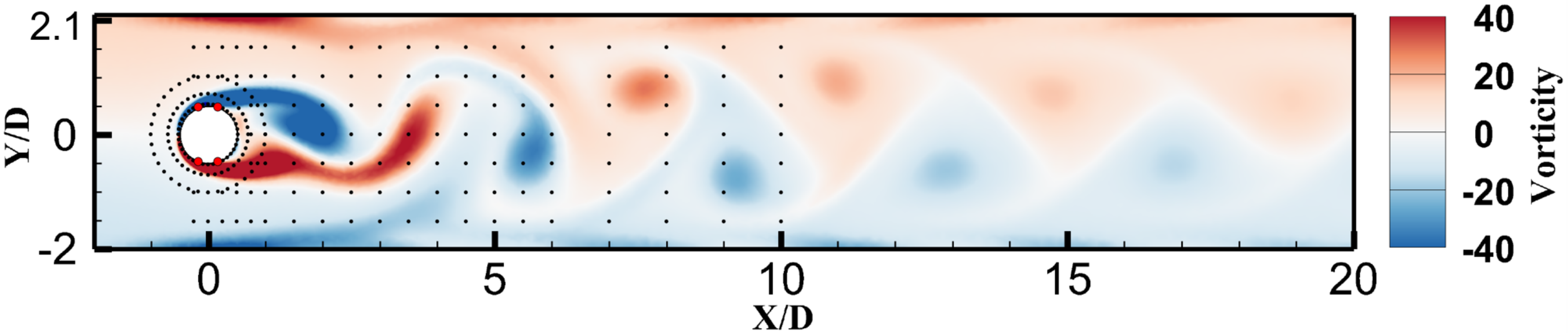}
  \caption{Unsteady non-dimensional vorticity wake behind the cylinder after
  flow initialization without active control. The location of the probes is
  indicated by the black dots. The location of the control jets is indicated by
  the red dots. This illustrates the configuration used to perform AFC for flow
  control past the circular cylinder.}
  \label{one_cylinder}
\end{figure*}

A simplified overview of the DRL framework used in the present study is
schematically depicted in Fig.\ \ref{RL_framework}. The framework can be divided
into two main parts: the environment and the learning agent. In the present
work, the former is the direct numerical simulation (DNS) for the flow past a
circular cylinder at low Reynolds number, as previously described. The latter
corresponds to a concrete deep reinforcement learning algorithm, proximal policy
optimization, which is described in detail later in this section. As illustrated
in Fig.\ \ref{RL_framework}, the learning agent interacts with the environment
through three channels: the state of the environment, the action chosen by agent
to influence the environment, and the reward signal that defines the goal of the
reinforcement learning problem. Specifically, the state is a partial observation
of the flow field. More concretely, the local value of the flow field sampled at
236 probes located around the cylinder and in its wake (black points in Fig.\
\ref{one_cylinder}) acts as the input based on which the agent can infer the
different flow features. These probes do not influence the flow field since the
extraction of local physical quantities of flow variables is carried out after
the numerical simulation ends at each time step. The ANN used by the agent to
parametrize the decision policy distribution is a 2-layer fully-connected
network with 512 neurons in each layer. The resulting action value provided by
the agent is then connected to the mass flow rate applied to each jet. The
reward function is the time-averaged drag of a training action penalized by the
absolute magnitude of the time-averaged lift, which can be expressed as follow:

\begin{equation}
  R_T=|F_D|_T-\beta|F_L|_T,
  \label{Eq:reward}
\end{equation}

\noindent where $|\cdot|_T$ indicates the average over an action time step
$T=100\delta t$ (see later), and $\beta$ is a parameter set to $0.2$ in the
present work. The lift penalization is used to avoid a ``cheating'' strategy
in which the jets could blow consistently in the same direction with maximum
strength after a given point in time. More details on the motivation for the penalization term
are discussed in the work of Rabault \textit{et al.} \cite{Rabault2019b}. In
general, a learning agent is able to use the state of the environment it
controls to take actions so as to optimize the cumulative value of the reward
function, which corresponds to the lowest drag.

The reinforcement learning algorithm used for training the ANN, known as
proximal policy optimization (PPO), is one of the
state-of-art reinforcement learning approaches and has been widely applied to
control tasks \cite{Rabault2019a, 8798254}. Compared with other DRL algorithms,
PPO is simpler to implement and tune while obtaining comparably good
performance. As the PPO algorithm has already been used in a variety of Fluid
Mechanics works, the reader interested in more details on the PPO algorithm
itself is invited to consult previous work on the topic \cite{Rabault2019a}. The
PPO method is episode-based, which means that the interactions between the agent
and the environment are broken into a number of training interaction sequences
\cite{Sutton2018}. The initial states for the training episodes at each Re are
first obtained by performing the simulation without active control until a
fully-developed unsteady wake, i.e., the K\'arm\'an vortex street, is observed.
The corresponding solution is stored and used as a starting point for
subsequent learning episodes. For the environment with four flow configurations,
the initial state is selected randomly from the initialized fields corresponding
to Re 100, 200, 300 and 400.

One possible discussion could be whether 236 probes are enough for the ANN to
have detailed information about the flow features and perform good or even
optimal control of the system. More generally, assessing the efficiency of the
decision made by partial observability of the system is a well-known difficulty
in reinforcement learning and remains an active and increasingly important
research challenge \cite{Recht2019}. Based on previous work on the topic
\cite{Rabault2019a} and our experience following preliminary tests during the
present study, 236 probes are found to be enough for the ANN to perform
adequate training and to attain satisfactory control performance. Much fewer
probes (less than 10) could also help the agent to learn a valid strategies,
but it will impair the control effects \cite{Rabault2019a}, i.e., lesser drag
reduction will be obtained compared to the results using more probes. With the 236 probes used
in present study, the agent is able to gain extensive information about the flow
configuration around the cylinder and its far-wake, which is important for taking
optimal actions. These probes are purely passive, and simply report the local
properties of the flow to the PPO algorithm, without influencing the flow.

In order to use the PPO algorithm on the present problem, two techniques are
implemented for structuring the interactions between the agent and the flow
environment. First, during the simulation, the action provided by the PPO agent
is updated only 200 times per episode, and is kept constant for a duration of
100 numerical simulation time steps (this defines the length of one action time step, i.e.,
the $T$ in Eq. \ref{Eq:reward}), corresponding to approximately 3.3\% of the
vortex shedding period. This limitation is added following the suggestion of
Rabault \textit{et al.} \cite{Rabault2019b}, and the necessity for such tuning
of the action frequency update has also been observed by Braylan \textit{et al.}
\cite{braylan:aaai15ws}. As a consequence, in the following, the difference will
be distinguished between the numerical timestep and the period at which action
update is applied. Second, the instantaneous mass flow rates obtained from the
actions are made continuous at the time scale of the numerical simulation $dt$
in order to avoid invalid physical jumps on pressure or velocity distribution
around the cylinder wall. Thus, the control value effectively applied changes
smoothly with time.

It should be emphasized that a balance needs to be found to avoid a too long
update interval which makes it impossible for the learning agent to respond to
the system fast enough, or too short update interval which means that the time
over which the action is applied is too short to observe a measurable effect on
the system, therefore making learning impossible during the first stage of the
training \cite{alex2018adaptive}. Furthermore, a constraint,
$|Q^*_i=Q_i/Q_{ref}|\leq 0.05$, is imposed for preventing non-physically large
actuations, where $Q_i$ is the mass flow rate of the $i-th$ jet and $Q_{ref}$ is the
reference mass flow rate intercepting the cylinder. This allows to avoid
divergence of the numerical simulation.

\section{Results and discussion}\label{sec:results}

\subsection{Active control for flow at higher Reynolds number}

Previous works \cite{Rabault2019a, Rabault2019b} have shown that ANNs trained by
DRL are capable of finding a good control strategy for controlling the flow
obtained in the present configuration at $Re=100$. However, it is known that the
Reynolds number has a strong influence on the complexity of such flows, how
chaotic the cylinder wake is, and ultimately laminar-to-turbulent transition of
the flow past a circular cylinder. For the present flow configuration, the wake
becomes more irregular at larger $Re$.

On the other hand, Protas and Wesfreid \cite{Protas2002} have proposed that two
parts contribute to the mean drag coefficient $C_D$ observed in such flows: one
is the drag $C_D^{base}$ of the steady and symmetric flow, and the other is the
drag $C_D^0$ resulting from the effect of vortex shedding:

\begin{equation}
  C_D=C_D^{base}+C_D^0.
\end{equation}

In other words, the averaged drag consists of contributions of steady and
unsteady parts, respectively. According to argument of Bergmann \textit{et al.}
\cite{Bergmann2005}, only the second part (due to oscillatory flow) can be
altered by AFC. Therefore, this provides an estimate of the optimal AFC drag
reduction attainable.

\begin{figure*}
  \centering
  \subfigure[]{
  \includegraphics[width=6.5cm, height=5.5cm]{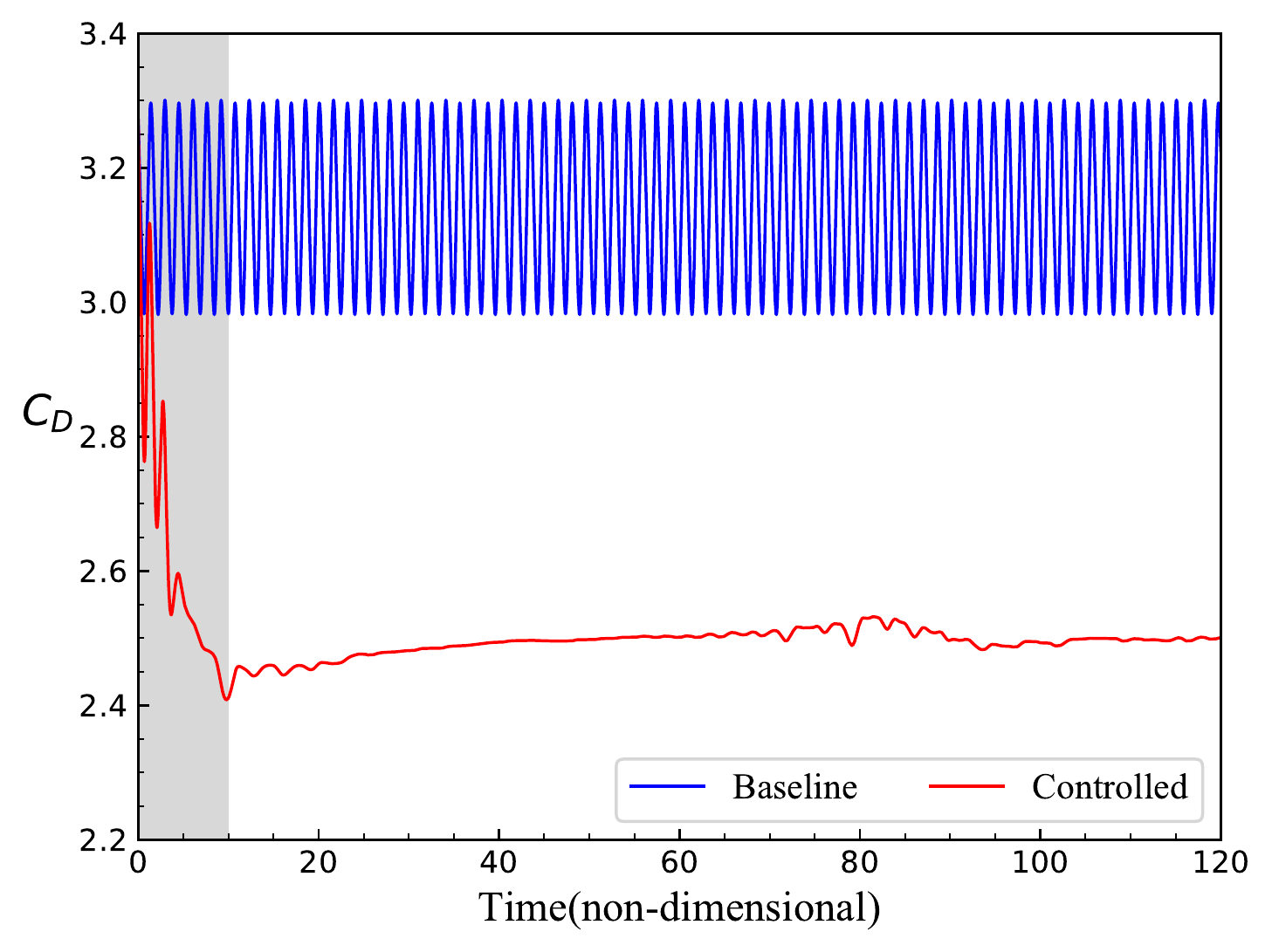}
  }
  \subfigure[]{
  \includegraphics[width=6.5cm, height=5.5cm]{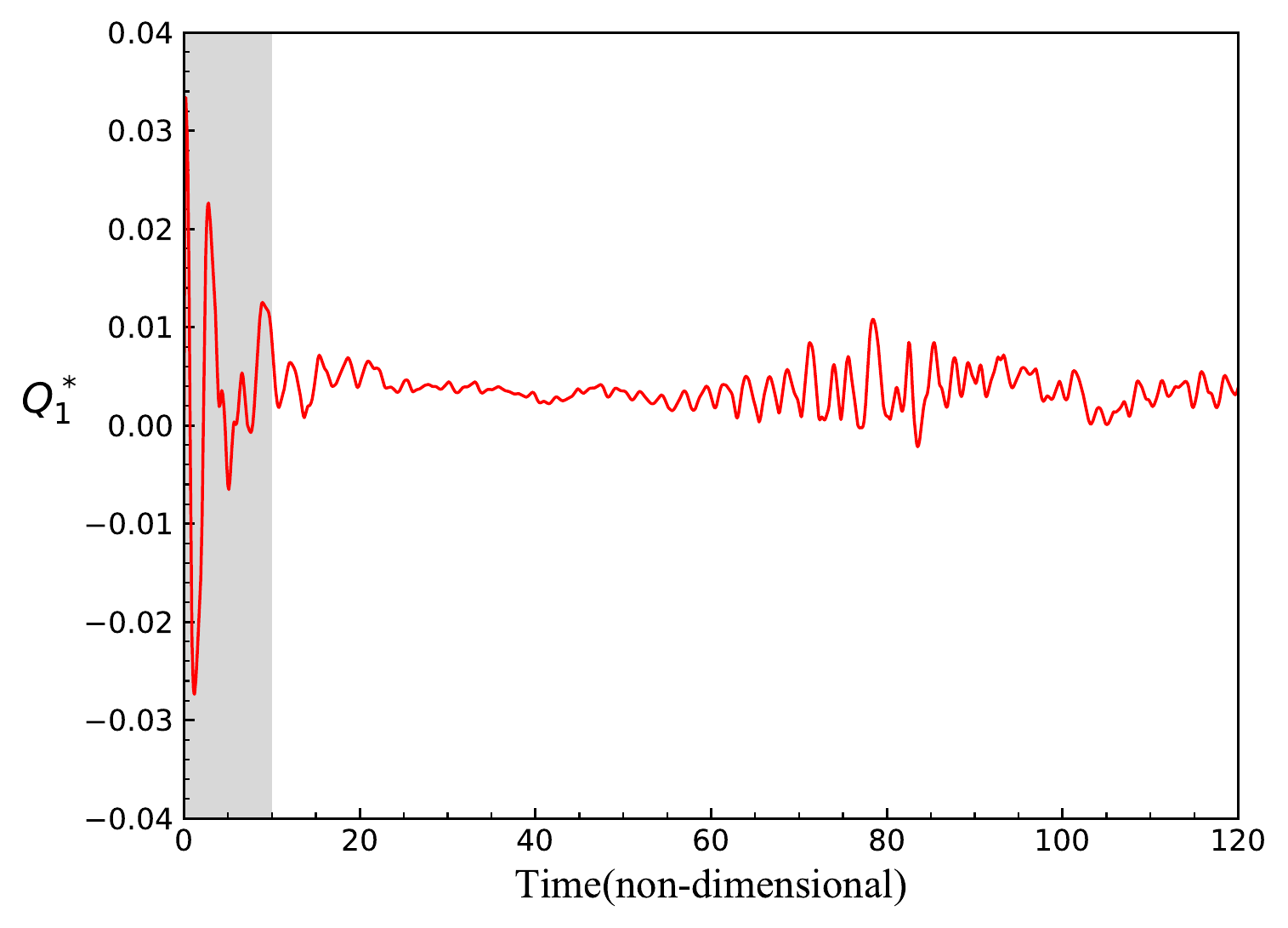}
  } \caption{Active control for flow at $Re=200$. (a) Time-resolved value of the
  drag coefficient $C_D$ with (controlled curve) and without (baseline curve)
  active flow control. (b) Time-resolved value of the normalized mass flow rate
  of one jet. It can be seen that the PPO agent found a good control strategy to
  attain a drag reduction of approximately 20.4\%. Two successive phases can be
  observed with control: in the first, relatively large actuations are performed
  to greatly reduce the drag, followed by a pseudo-periodic regime in which only
  small control actuations is needed.}
\label{Re_200}
\end{figure*}

\begin{figure*}
  \centering
  \subfigure[]{
  \includegraphics[width=6.5cm, height=5.5cm]{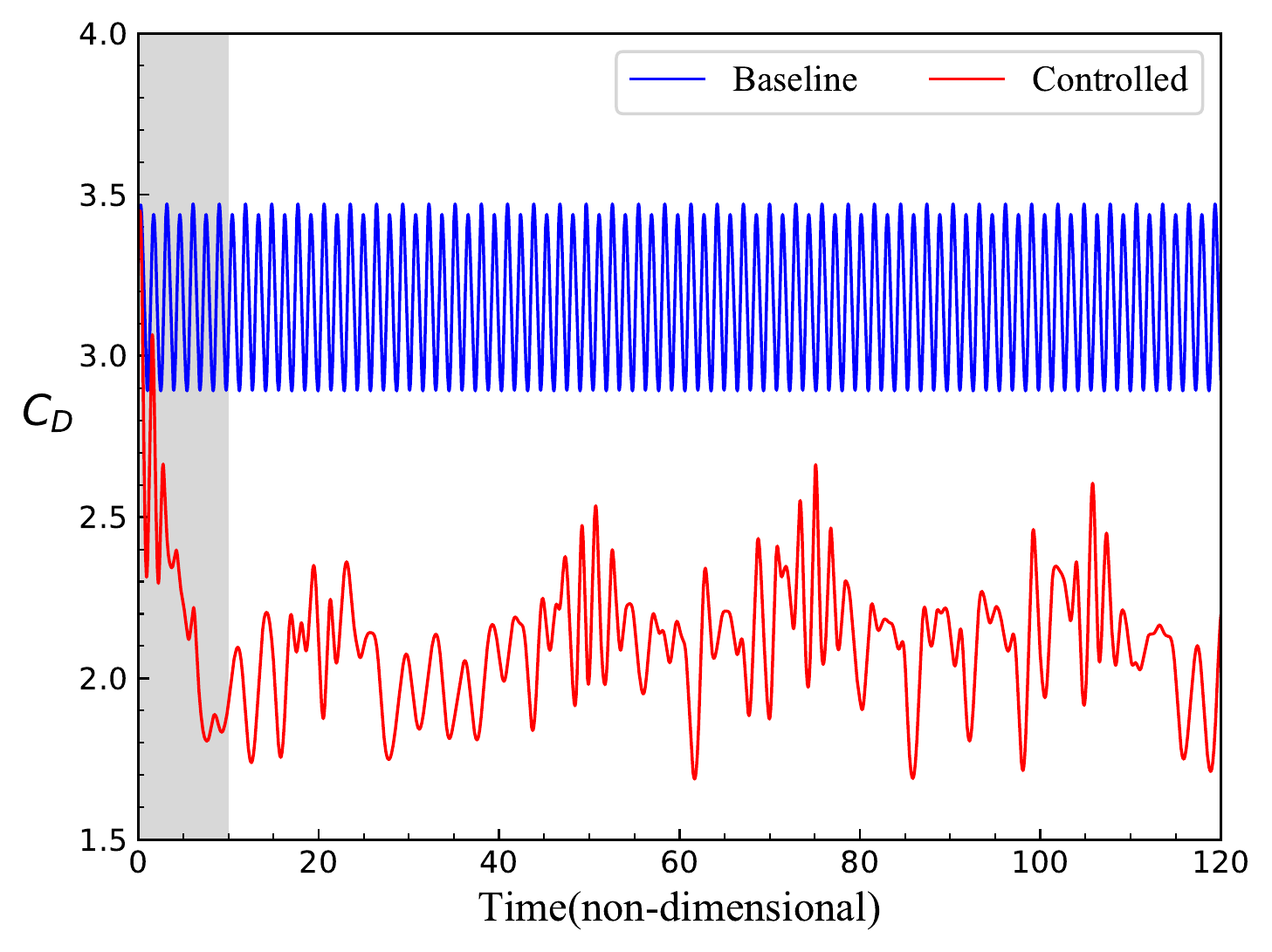}
  }
  \subfigure[]{
  \includegraphics[width=6.5cm, height=5.5cm]{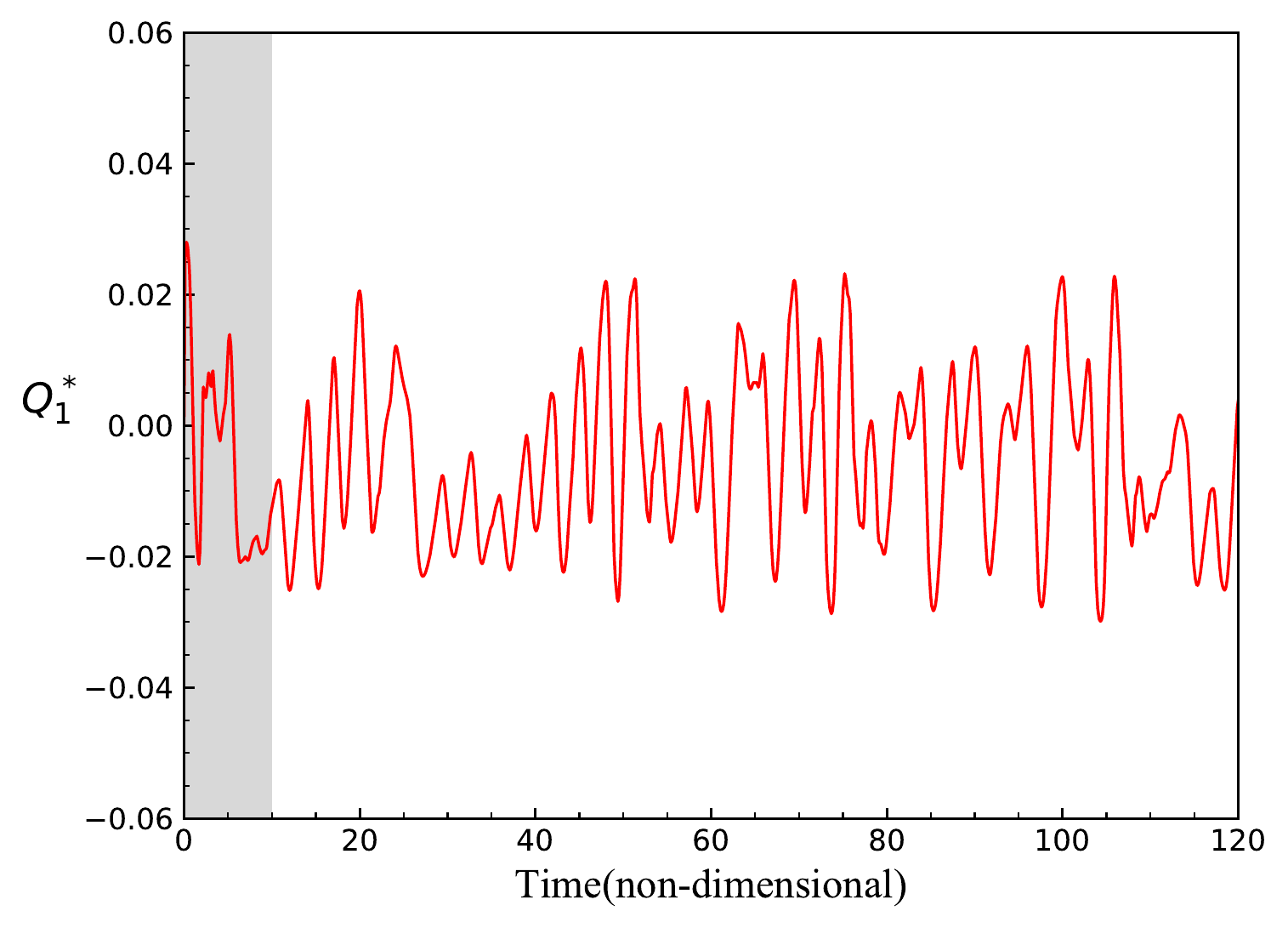}
  } \caption{Active control for flow at $Re=400$. (a) Time-resolved value of the
  drag coefficient $C_D$ with (controlled curve) and without (baseline curve)
  active flow control. (b) Time-resolved value of the normalized mass flow rate
  of one jet. Two successive phases can be distinguished. Similar to the flow
  with control at $Re=100$ (cf. the work of Rabault \textit{et al.}
  \cite{Rabault2019a}) and 200, a clear reduction of drag of approximately
  33.1\% is obtained in the first phase. However, in contrast to Re 100 and 200,
  no large decrease of actuations is observed in the second phase. This is due
  to the inherent instability of the flow at larger Reynolds numbers, and
  illustrates the ability of the PPO algorithm to control systems with
  pseudo-chaotic properties.}
\label{Re_400}
\end{figure*}

Since it has been demonstrated that the contribution of $C_D^0$ increases with
$Re$ \cite{Bergmann2005}, it is natural to investigate the control performance
of ANNs trained through DRL for flow for increasing $Re$. Consequently, two
individual ANNs are trained to obtain control strategies for flow with $Re=200$
and $400$, respectively. Here, the same control configurations as previous works
\cite{Rabault2019a} are used, i.e., two jets located at the top and bottom
extremities of the cylinder. The drag coefficients when control is applied by
the ANNs after training are shown in Fig.\ \ref{Re_200} and Fig.\ \ref{Re_400},
with the results of baseline flow (i.e. without control) being shown as a
reference. The drag reduction is calculated as
$(|C_D|_{base}-|C_D|_{control})/|C_D|_{base}$, where $|C_D|_{base}$ and
$|C_D|_{control}$ are the mean value for drag coefficients $C_D$ in the case
without and with active control, respectively. A drag reduction of
approximately 20.4\% is observed at $Re=200$ and the final control result is
satisfactory, though small oscillations still exist. By contrast, at $Re=400$,
although a reduction of approximately 33.1\% for the averaged drag was achieved,
the nonlinear essence of the transitional flow makes it hard for the DRL agent
to find a fully stabilized control strategy and to completely suppress
oscillations in the drag coefficient. However, the amplitude of the drag
oscillations as well as their frequency is still decreased, implying that the
DRL agent indeed learns some strategy that allows effective control. Similar to
what has been observed for the same configuration at $Re=100$
\cite{Rabault2019a}, the active flow control consists of two successive phases.
In the first phase (non-dimensional time ranging from 0 to approximately 10), a
clear drag reduction is achieved by performing relatively large actuations. The
flow is then modified into a pseudo-periodic regime where smaller actuations are
used at $Re=200$. For the flow at $Re=400$, however, in the second phase, there
is less attenuation of the actuations, resulting in big oscillations of drag
coefficients even with control. Therefore, it appears that the flow in
transitional regimes is quite unstable, which easily leads to a collapse of the
modified flow configuration, and in turn calls for large actuations to regain
control of the system. This illustrates the ability of the PPO algorithm to
perform control of pseudo-chaotic systems such as obtained from the simulation
of flows at moderate to high Reynolds numbers, in good agreement with previously
published results \cite{Rabault2019a, Rabault2019b}.

\subsection{Effect of smoothing interpolation functions}

As explained in Sec.\ \ref{sec:DRL}, it is of great importance to use suitable
methods to interpolate the intrinsically time-discretized output of the ANN to
continuous systems. This is still a topic of ongoing research with no
clear optimal solution \cite{Sutton2018}. The present work chooses to
directly interpolate between action updates to generate the control value at
each simulation time step. This is simple to implement while maintaining a good
performance for policy training and action selection. The interpolation must
follow some principles such as smoothness and continuity to avoid numerical
instability caused by non-physical phenomenon such as pressure jump in the fluid
flow.

The interpolation can be performed in several fashions by considering the
different relationships among action updates. Rabault \textit{et al.}
\cite{Rabault2019b} proposed an exponential decay law based on the control value
from the previous action. More precisely, they use the following equation with
$\alpha=0.1$ to calculate new control value: 

\begin{equation}
  c_{i+1}=c_i+\alpha(a_j-c_i),
  \label{smooth1}
\end{equation}

\noindent where $c_i$ is the control value at previous numerical time step,
$c_{i+1}$ is the new control, $a_j$ is the action updated by the ANN. Note that
the subscript $i$ means $i-th$ numerical time step, which is connected to the
time step $dt$ of the simulation. By contrast, the subscript $j$ indicates
$j-th$ action update interval, which corresponds to the number of the action
update during an episode, and happens at a period $T = 100 \delta$.

\begin{figure*}
  \centering
  \includegraphics[width=16cm, height=6.5cm]{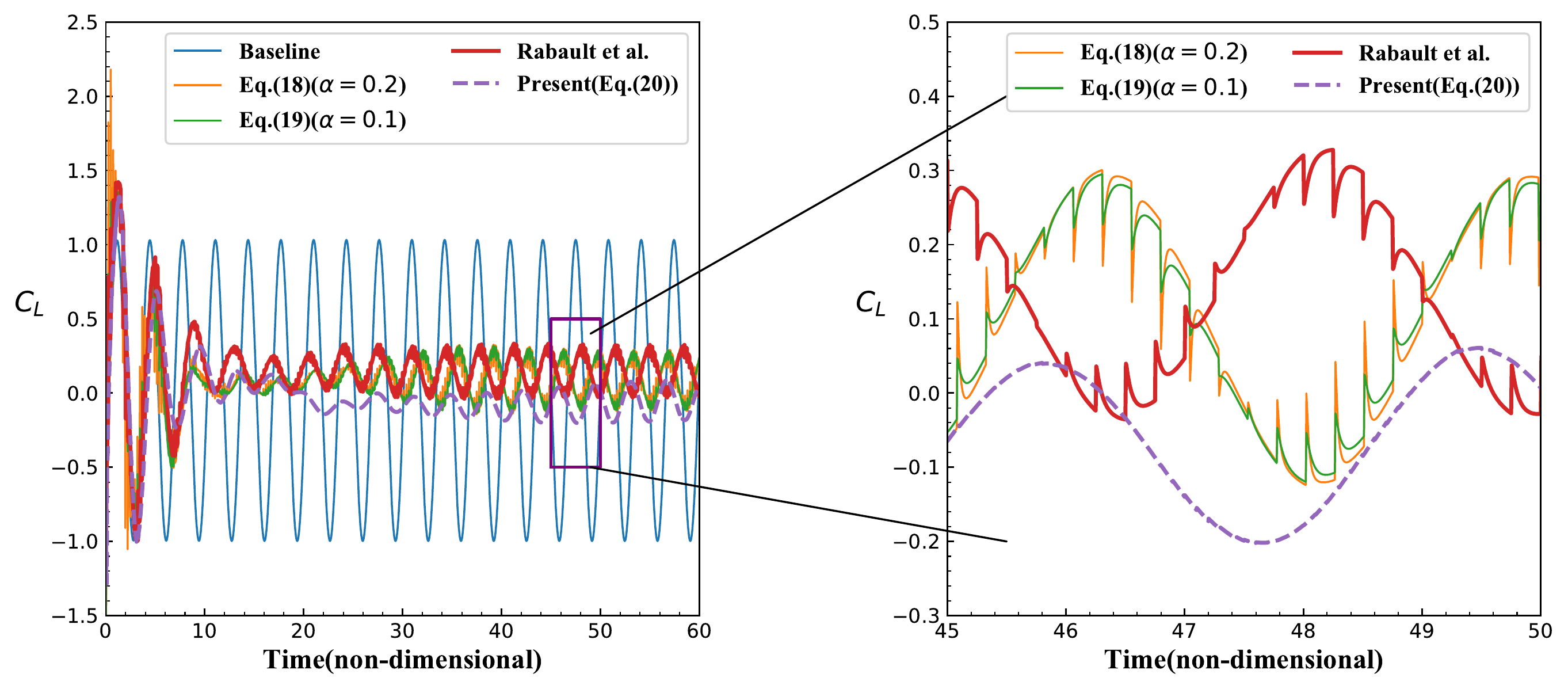}
  \caption{Comparison of time-resolved value of lift coefficients $C_L$ at
    $Re=100$ with active control trained using Eq.\ (\ref{smooth1}) (used by
    Rabault \textit{et al.} \cite{Rabault2019a} with $\alpha=0.1$), Eq.\
    (\ref{smooth2}), Eq.\ (\ref{smooth3}) and Eq.\ (\ref{smooth4}) (used in the
    present work), respectively. The linear smooth law, i.e., Eq.\
    (\ref{smooth4}), shows best performance as jumps in lift are almost
    completely suppressed.}
  \label{smooth_test}
\end{figure*}

The strategy obtained using Eq.\ (\ref{smooth1}) is able to stabilize the vortex
alley and to reduce drag by approximately 8\% at $Re=100$. The exponential decay
law performs well for the convergence of the control values, however, there are
distinct problematic jumps in lift, indicating that the flow state with control
is still not perfectly stable. This is visible in Fig.\ \ref{smooth_test}. Some
other schemes for interpolation also show similar problems. For example, one can
consider more previous control value for performing an update, or use a
nonlinear interpolation, which can be implemented respectively as:

\begin{equation}
  c_{i+1}=c_i+\alpha(a_j-c_i)+\alpha(a_{j-1}-c_{i-1}),
  \label{smooth2}
\end{equation}

\begin{equation}
  c_{i+1}=c_i+\alpha(a_j-c_i)+\alpha(a_j-c_i)^2.
  \label{smooth3}
\end{equation}

After extensive trial-and-error, it is finally found that linear interpolation
between two actions (corresponding to Eq.\ (\ref{smooth4}), see under) shows a
comparable performance to exponential decay law while effectively eliminating
the oscillations of the lift coefficient. Fig.\ \ref{smooth_test} shows a
comparison of the control effects with the different smoothing laws discussed
above. Obviously, Eq.\ (\ref{smooth4}) shows best control performance. The
corresponding interpolation law is defined as:

\begin{equation}
  c_{i}=a_{j-1}+\frac{a_j-a_{j-1}}{N_e},
  \label{smooth4}
\end{equation}

\noindent where $N_e$ is the number of numerical time steps between two
consecutive updates of actions, and $n=1, 2, \dots, N_e$ is the current control
step.

\subsection{Training a model over a range of Reynolds numbers}

To validate the versatility of an artificial neural network trained by deep
reinforcement learning to control a flow across different Reynolds numbers, a
learning environment supporting four flow configurations with $Re$ varying
within the discrete set 100, 200, 300 to 400 is used to train a single ANN.
Therefore, the aim here is to train one ANN to perform effective control over a
range of flow parameters, in a robust fashion. In this case, four jets are
located on the upper and lower sides of the cylinder, as described in Fig.\
\ref{BC}. Due to the learning process being treated on an episode base, each
flow simulation is first run without active control until a fully-developed
unsteady wake, i.e., the K\'arm\'an vortex street, is observed, and the
corresponding state is dumped and selected randomly as an initial start state
for subsequent learning episodes. Here, the multi-environment approach proposed
by Rabault \textit{et al.} \cite{Rabault2019b} is adapted, and the probability
for every flow state to be selected as the initial state of an episode is equal.
Since every environment is independent of the others, that is, episodes do not
influence each other due to the use of distinct initialization fields at
distinct Reynolds numbers \cite{Sutton2018}, the agent has to remember features
for different flow configurations so that the knowledge learned by the ANN for
one flow will not be altered by training on others.

\begin{figure*}
  \centering
  \subfigure[]{
    \includegraphics[width=7.5cm, height=6.5cm]{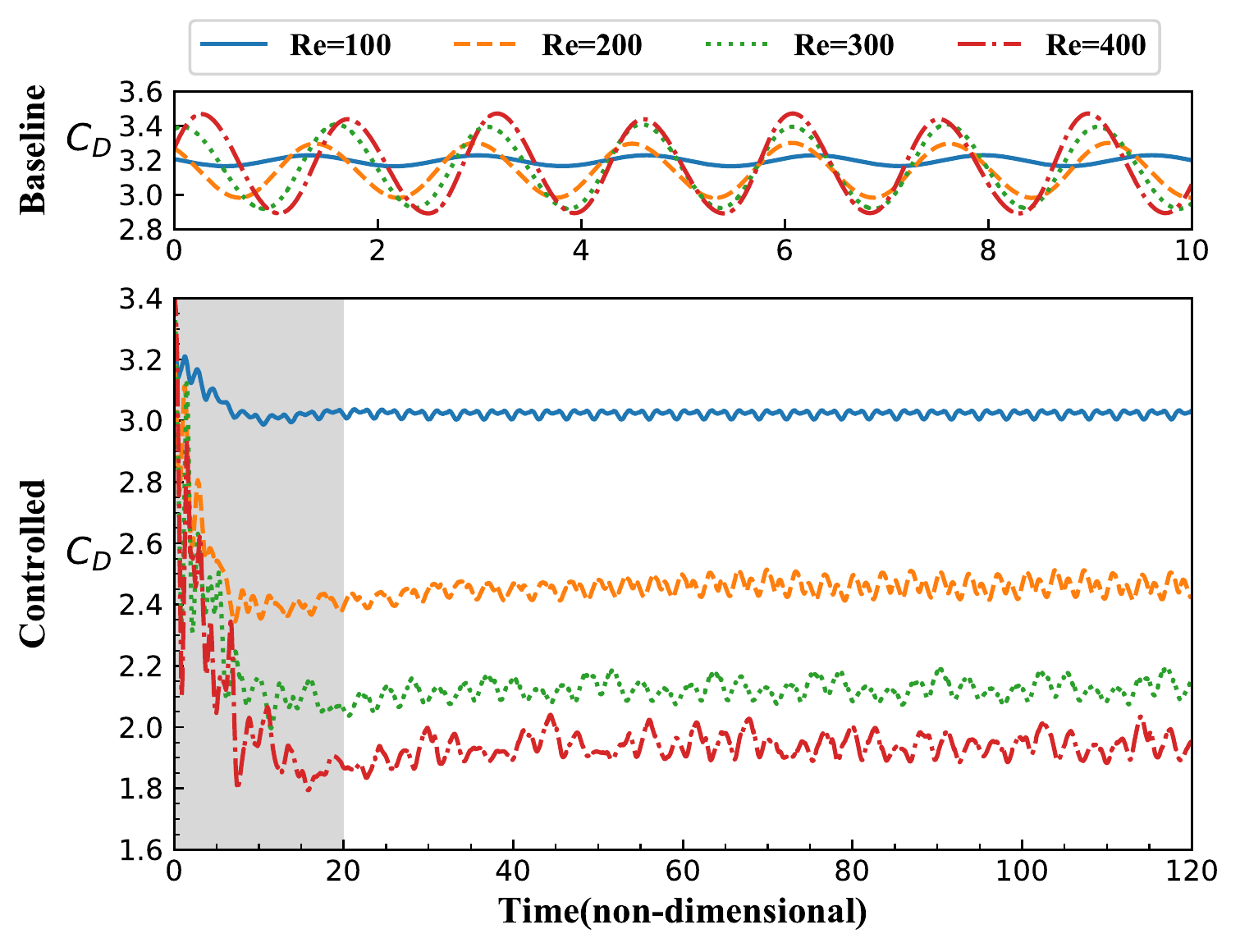}
  }
  \subfigure[]{
    \includegraphics[width=7.5cm, height=6.5cm]{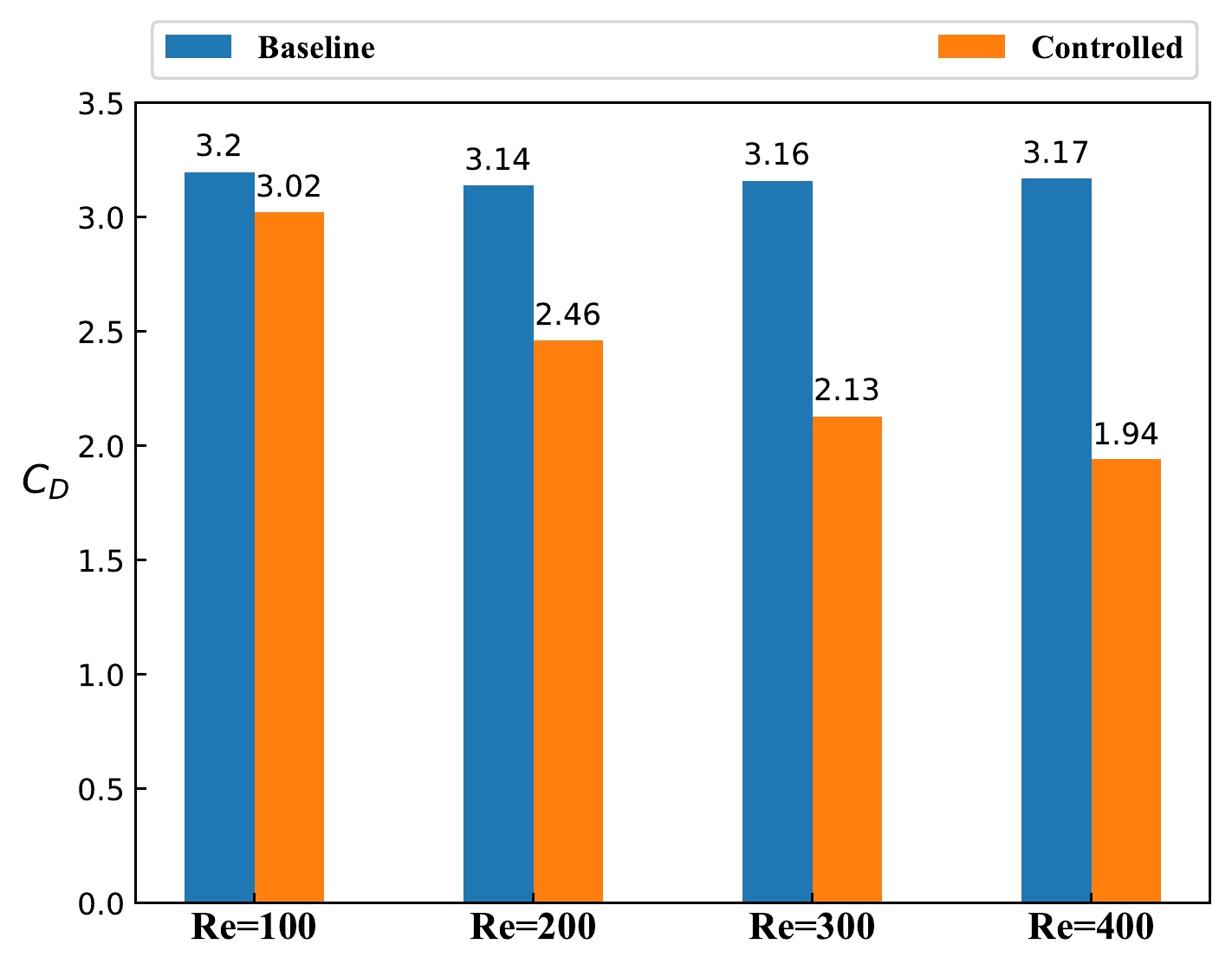}
  } \caption{Illustration of the control performance of the global agent
  (Controlled curves) for flows corresponding to  $Re=100$, 200, 300 and 400
  compared with the case without control (Baseline). (a) Time series of the drag
  coefficients $C_D$. (b) The average of the drag coefficients $C_D$. The drag
  is reduced by approximately 5.7\%, 21.6\%, 32.7\%, and 38.7\% when $Re$=100,
  200, 300, and 400 respectively. Similarly to what can be observed in Fig.\
  \ref{Re_200} and Fig.\ \ref{Re_400}, the active flow control consists of two
  successive stages. However, in comparison, the first stage of control takes a
  longer time (up to a non-dimensional time of approximately 20) compared with
  the case when the controlled strategy is tuned to a single Re value.}
  \label{general_Re_results}
\end{figure*}

The time series for the drag coefficients obtained using the global control
strategy after 800 episodes when $Re=100, 200, 300, 400$, are compared with
baseline flow (without active flow control), as shown in Fig.\
\ref{general_Re_results}. Compared with the results presented by Rabault
\textit{et al.} \cite{Rabault2019a} where the control strategy is discovered
through training in an environment consisting of one single flow configuration
($Re=100$), the global control strategy becomes slightly less effective at
$Re=100$, but the overall control strategy is significantly more robust since
the obtained ANN is able to adapt the actuation to perform near-optimal control
(see later in the text) at all $Res$ within the range 60-400. A drag reduction
of approximately 5.7\%, 21.6\%, 32.7\%, and 38.7\% is obtained when $Re$=100,
200, 300, and 400, respectively. Similar to the results presented in Fig.\
\ref{Re_200} and Fig.\ \ref{Re_400}, the process of active flow control is
composed of two phases. The main difference is that it takes a longer time (up
to a non-dimensional time of approximately 20) for attaining the typical value
of the drag reduction (i.e. the first phase of the control strategy takes a
longer time to complete). In addition, slightly larger fluctuations can be
observed during this phase, especially for higher $Re$.

One interesting result of this experiment is that the active control strategy
trained over a range of Reynolds numbers shows comparatively good performance
compared with the results shown in Fig.\ \ref{Re_400}. While the average
reduction of drag is close, the oscillations in drag are greatly suppressed with
the global control strategy. This may be due to two factors. First, only two
synthetic jets with angular coordinates $90^\circ$  and $270^\circ$ are used for
the results in Fig.\ \ref{Re_400}, while the global control strategy is allowed
to control 4 jets, therefore, allowing a more fine-grained control. Second, the
training of the global control strategy is performed over a range of Reynolds
numbers, therefore, presenting more variability during training. For further
exploration of this question, several independent training runs are launched
using same control configurations as described in Fig.\ \ref{one_cylinder}, i.e.
4 jets, for the flow with $Re=400$ (the learning environment is then composed of
one flow configuration, i.e. no global strategy is used). The drag coefficients
with control show in this case no big difference with what is presented in
Fig.\ \ref{Re_400} (the average of drag with control is similar and still
exhibits large oscillations). Such results prove the robustness and good
performance of the ANN obtained with global training, and points to the
utility of training the ANN over a range of conditions. On the other hand, it
also indicates that for much more complex systems, an efficient way to obtain
good control strategies may be to embed a number of similar but slightly
different systems inside the learning environment. This is in good agreement
with the commonly accepted concept of transfer learning (TL), the core idea of
which is that knowledge gained from one task can help the learning performance
in a similar but slightly different task, and improve overall performance
\cite{taylor2009transfer}.

\begin{figure*}
  \centering
  \subfigure[]{
  \includegraphics[width=7.5cm, height=6.4cm]{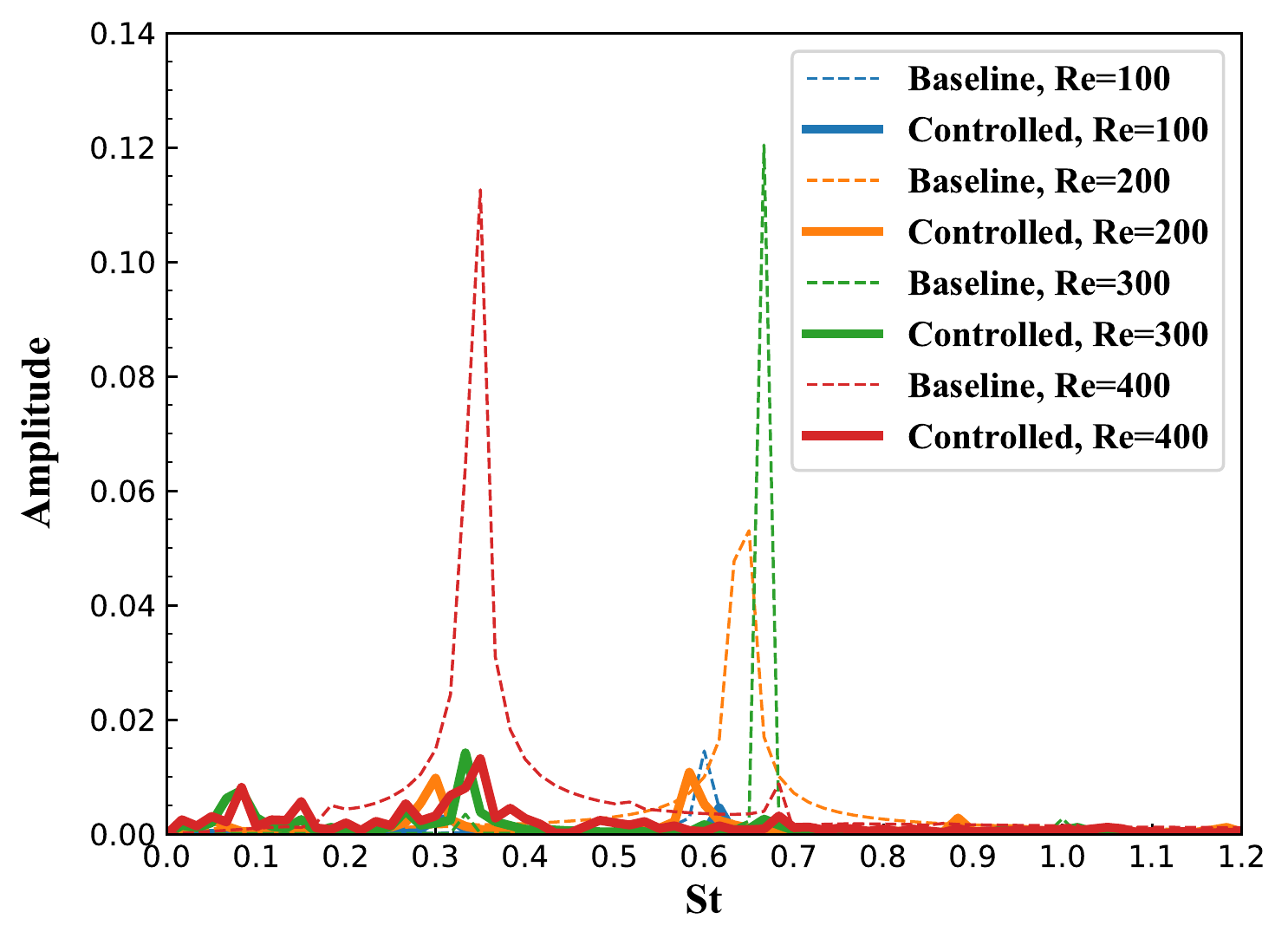}
  }
  \subfigure[]{
  \includegraphics[width=7.5cm, height=6.4cm]{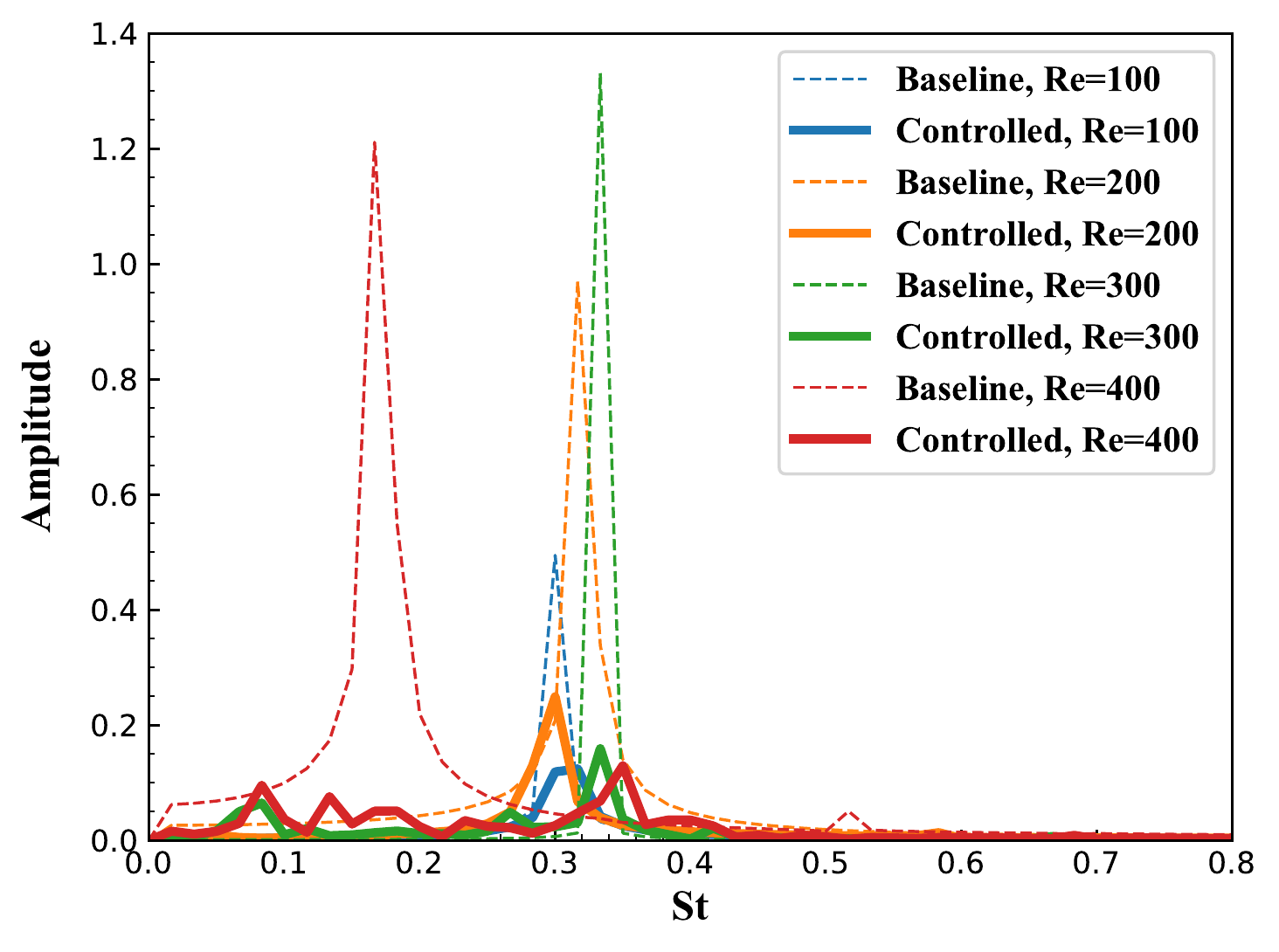}
  } \caption{FFT analysis of drag coefficients $C_D$ (a) and lift coefficients
  $C_L$ (b) subtracted by their mean values. The baseline curve corresponds to
  the flow without control while the controlled curves mean that the flow is
  controlled by the ANN. The control effects are clearly visible: the amplitudes
  corresponding to fluctuations of both drag and lift are greatly reduced and
  the characteristic frequencies of the flow fields are modified.}
  \label{fft_drag_lift}
\end{figure*}

\begin{figure*}
  \centering
  \subfigure[]{
  \includegraphics[width=8cm, height=3.5cm]{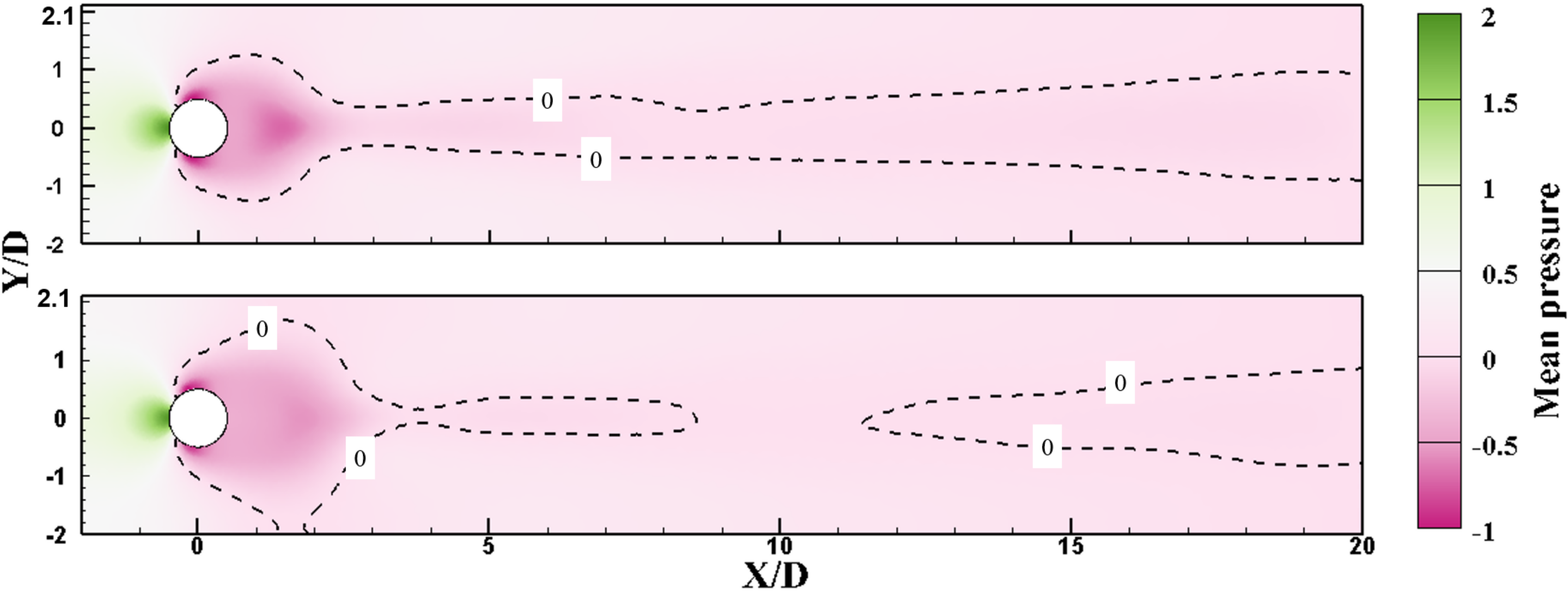}
  }
  \subfigure[]{
  \includegraphics[width=8cm, height=3.5cm]{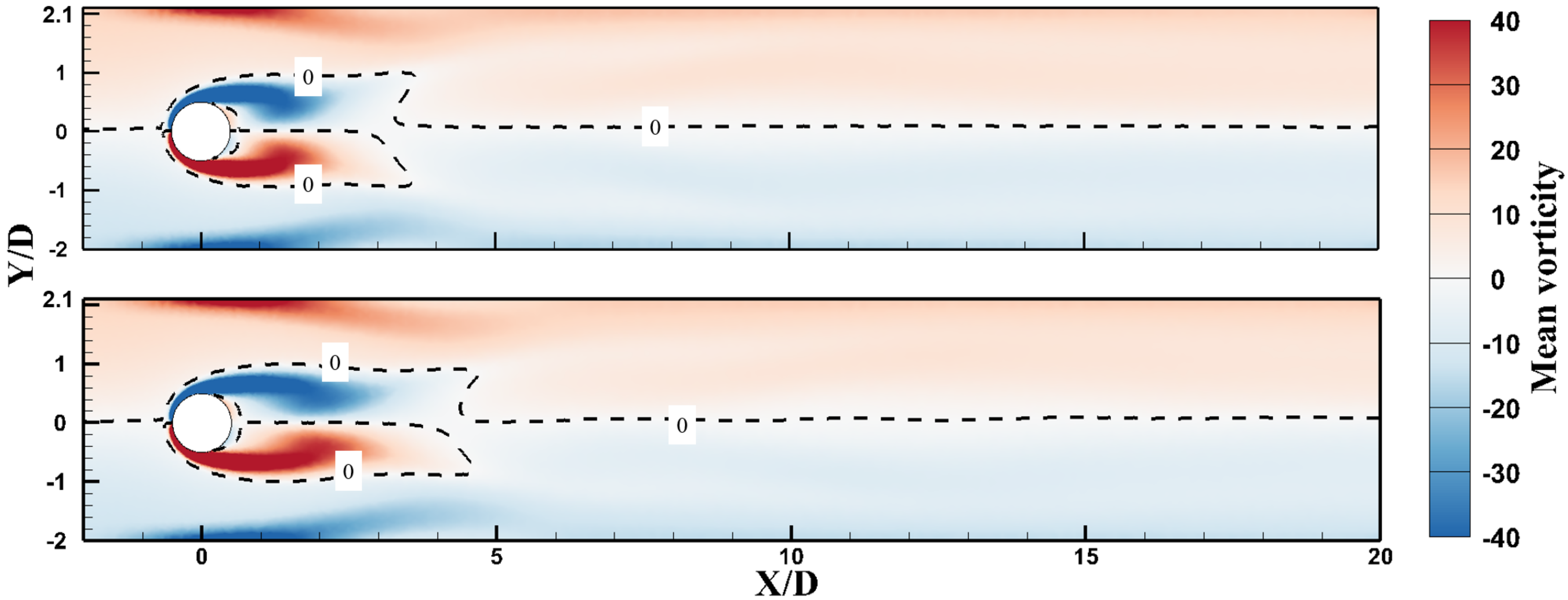}
  }
  \subfigure[]{
  \includegraphics[width=8cm, height=3.5cm]{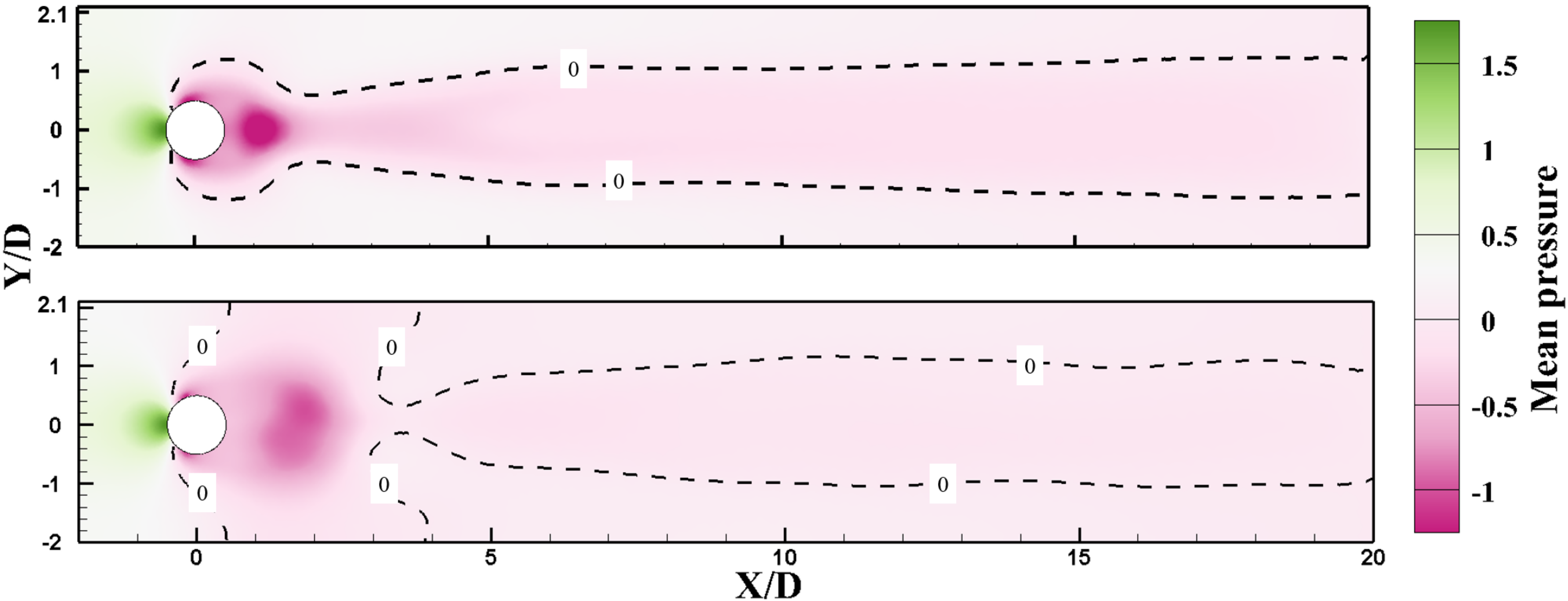}
  }
  \subfigure[]{
  \includegraphics[width=8cm, height=3.5cm]{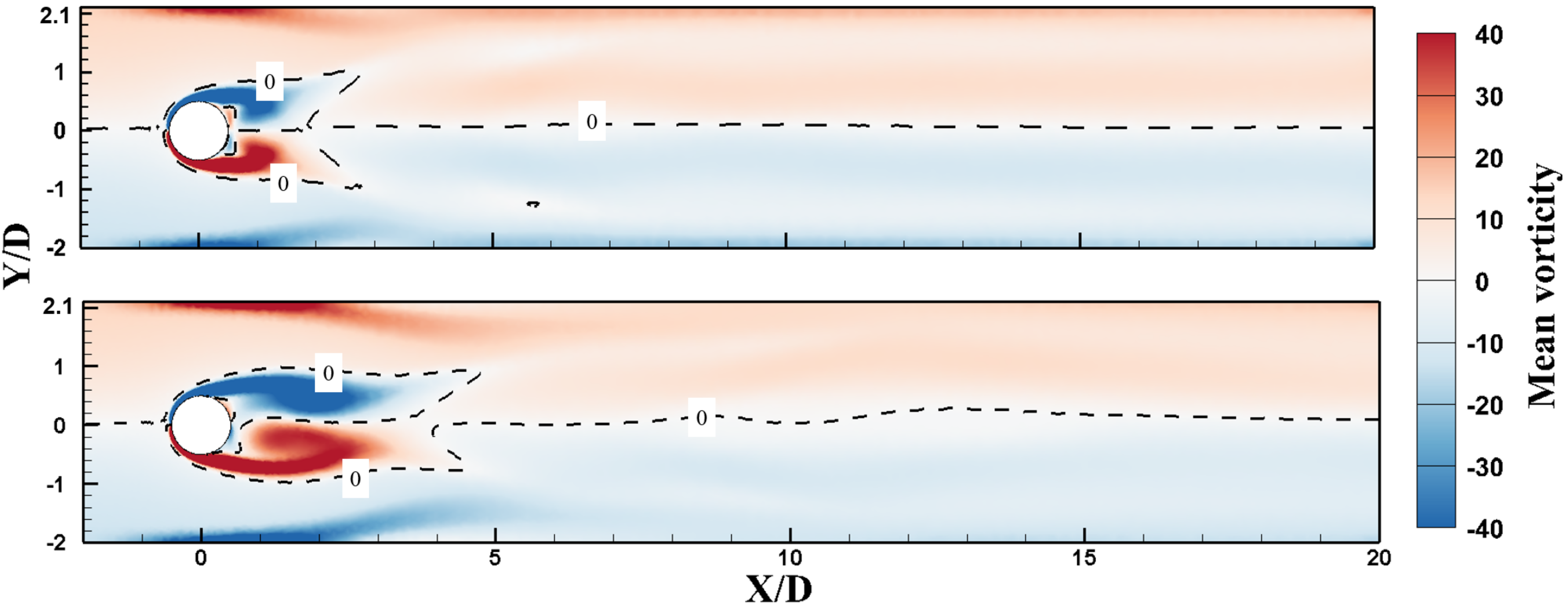}
  }
  \subfigure[]{
  \includegraphics[width=8cm, height=3.5cm]{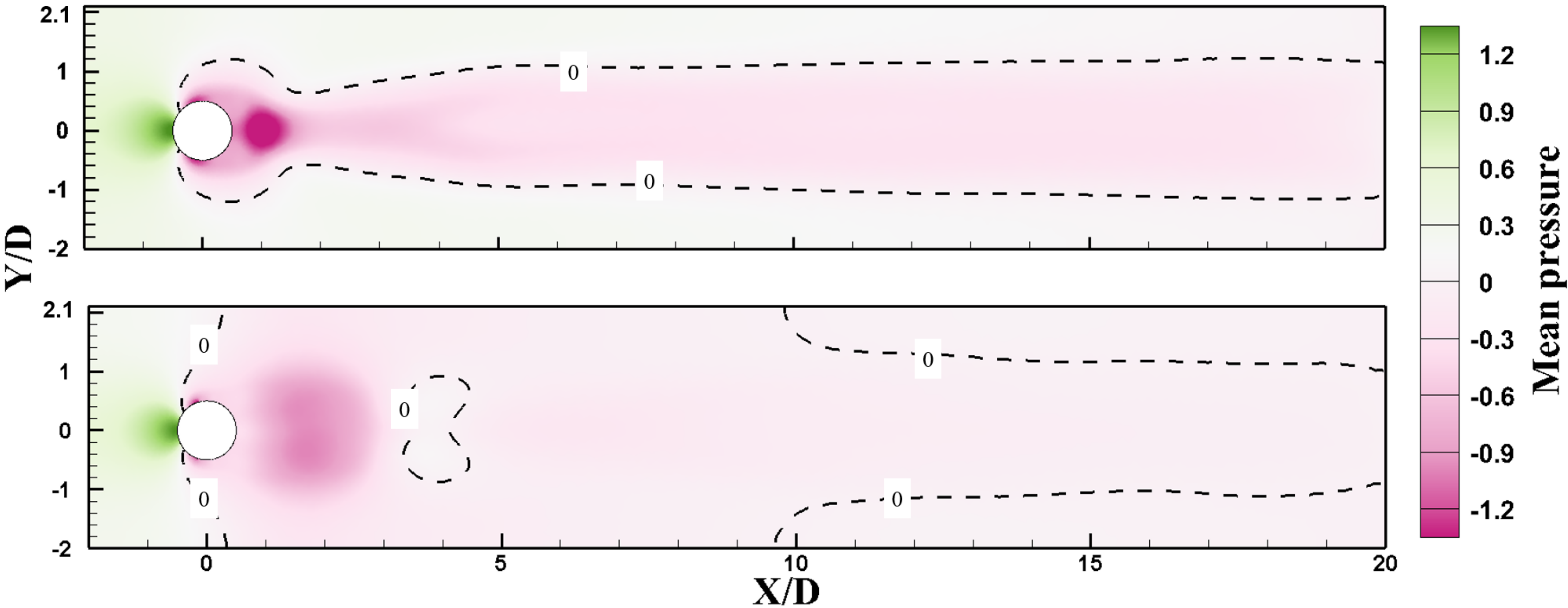}
   }
  \subfigure[]{
  \includegraphics[width=8cm, height=3.5cm]{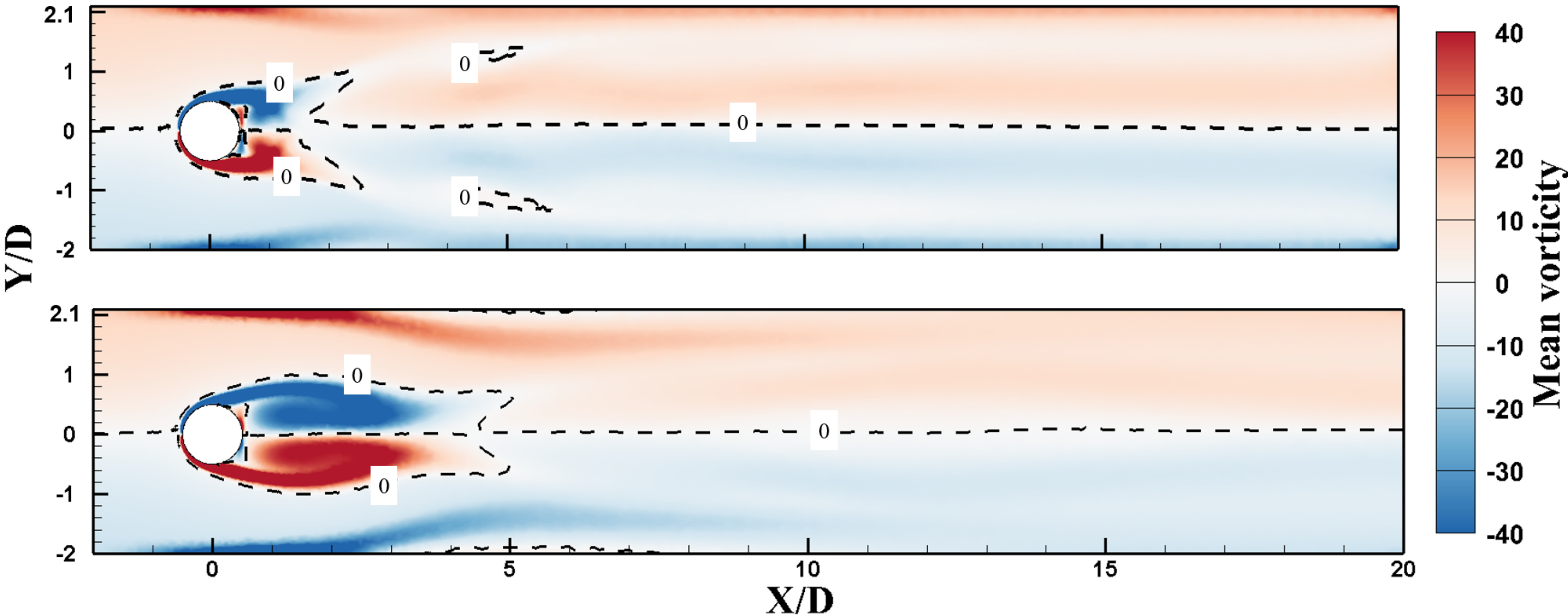}
  }
  \subfigure[]{
  \includegraphics[width=8cm, height=3.5cm]{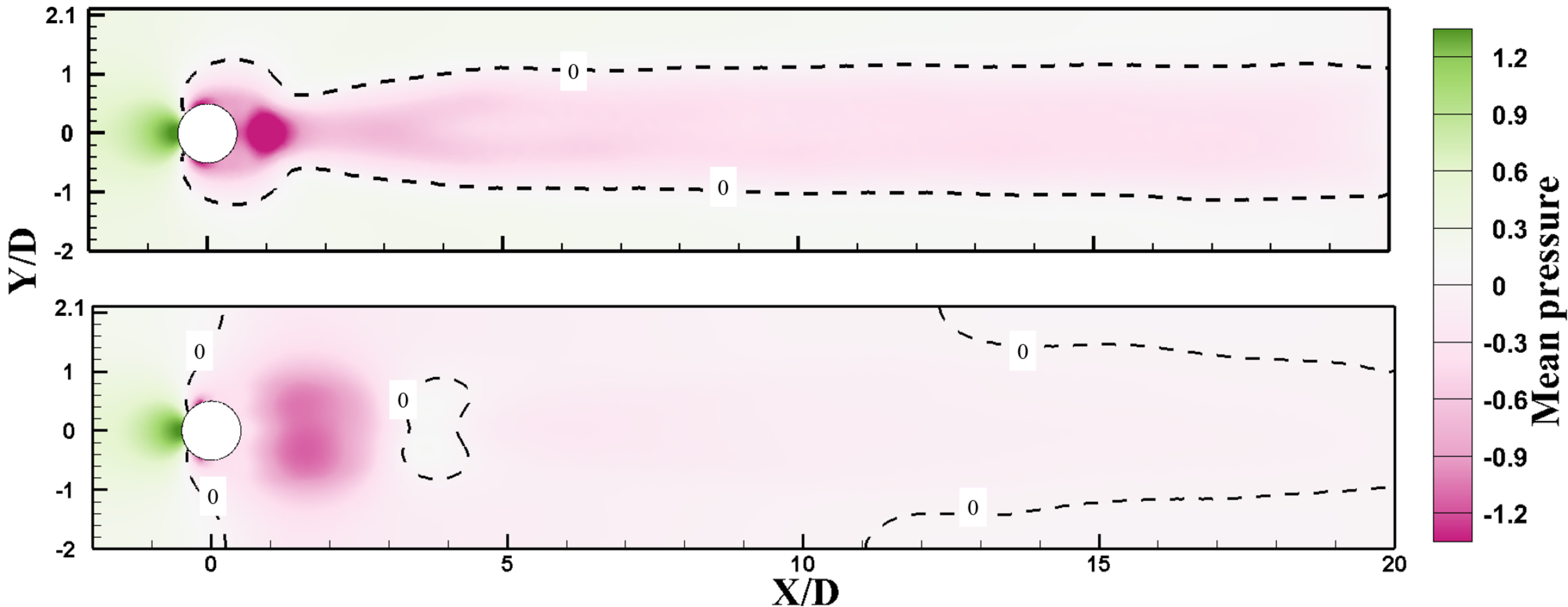}
  }
  \subfigure[]{
  \includegraphics[width=8cm, height=3.5cm]{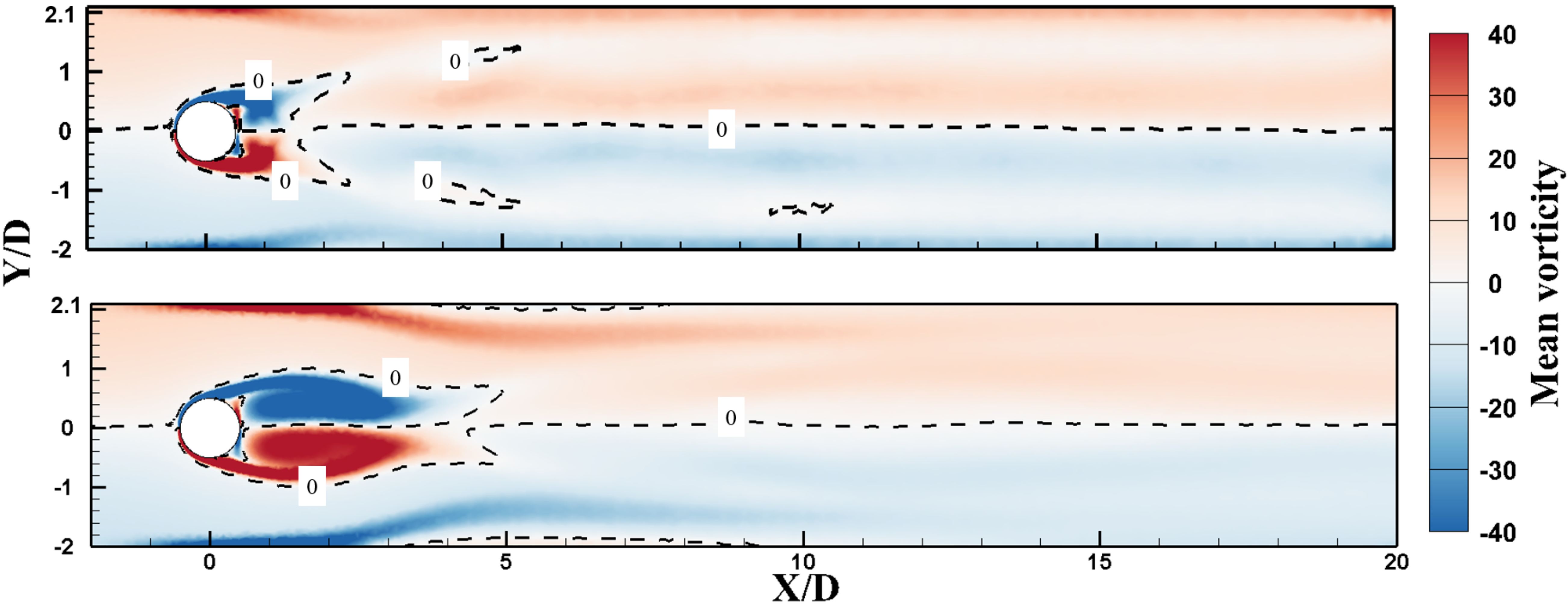}
  } \caption{Comparison of mean pressure (left) and vorticity (right) without
  (top part of each double panel) and with (bottom part of each double panel)
  active control. The Reynolds numbers for the four rows of double panels from
  top to bottom are 100 (a, b), 200 (c, d), 300 (e, f) and 400 (g, h),
  respectively. The colour bar is common to both parts of each double panel.
  When the active control is applied, the area of separated wake increases and
  the vortex shedding from the cylinder is substantially enlarged. The former
  flow morphology is associated with the reduction of drag and lift, while the
  later is connected to the lower oscillations in these two forces. For blunt
  bodies acting in the flow regime considered, the largest contribution to the
  drag coefficient is due to the pressure fall in the wake, and it is clearly
  visible at all Res that the control strategy found allows to mitigate this
  pressure drop immediately behind the cylinder.}
  \label{pressure_vorticity}
\end{figure*}

As expected, the control strategy is more effective at reducing drag for larger
$Re$, due to the relative increase of the controllable contribution of drag
previously discussed, i.e. $C_D^0$ is relatively bigger at higher Re. To further
analyze the results, the fast Fourier transformation (FFT) is applied for
investigating the frequency of drag and lift time series with and without active
control (60000 numerical time steps are used for calculating the FFT). For making the results more
easily visible, the drag and lift coefficients are subtracted by their average
value before FFT analysis is applied, so that the purely oscillatory properties
of the coefficients in question are revealed. As visible in Fig.\
\ref{fft_drag_lift}, there is an obvious reduction on the amplitude of drag
fluctuations. Moreover, the characteristic frequency of the flow system actively
controlled by the ANN is also modified. These results are similar to what was
described by Rabault \textit{et al.} \cite{Rabault2019a}.

To study in more details the effect of the control on the flow field, a visual
comparison of the flow undergoing control against the mean pressure and
vorticity of the uncontrolled flow is presented in Fig.\
\ref{pressure_vorticity}. As can be observed, the area of separated wake
increases when the active control is applied. Moreover, the vortex shedding from
the cylinder has been substantially enlarged and expanded by the synthetic jets,
which causes the observed reduced fluctuations. The resulting flow approaches the
state featuring symmetric characteristic as will be discussed next. As a
consequence, the pressure drop in the wake of the cylinder becomes lower,
causing the reduction of drag. 

\begin{figure*}
  \centering
  \includegraphics[width=11.5cm, height=10cm]{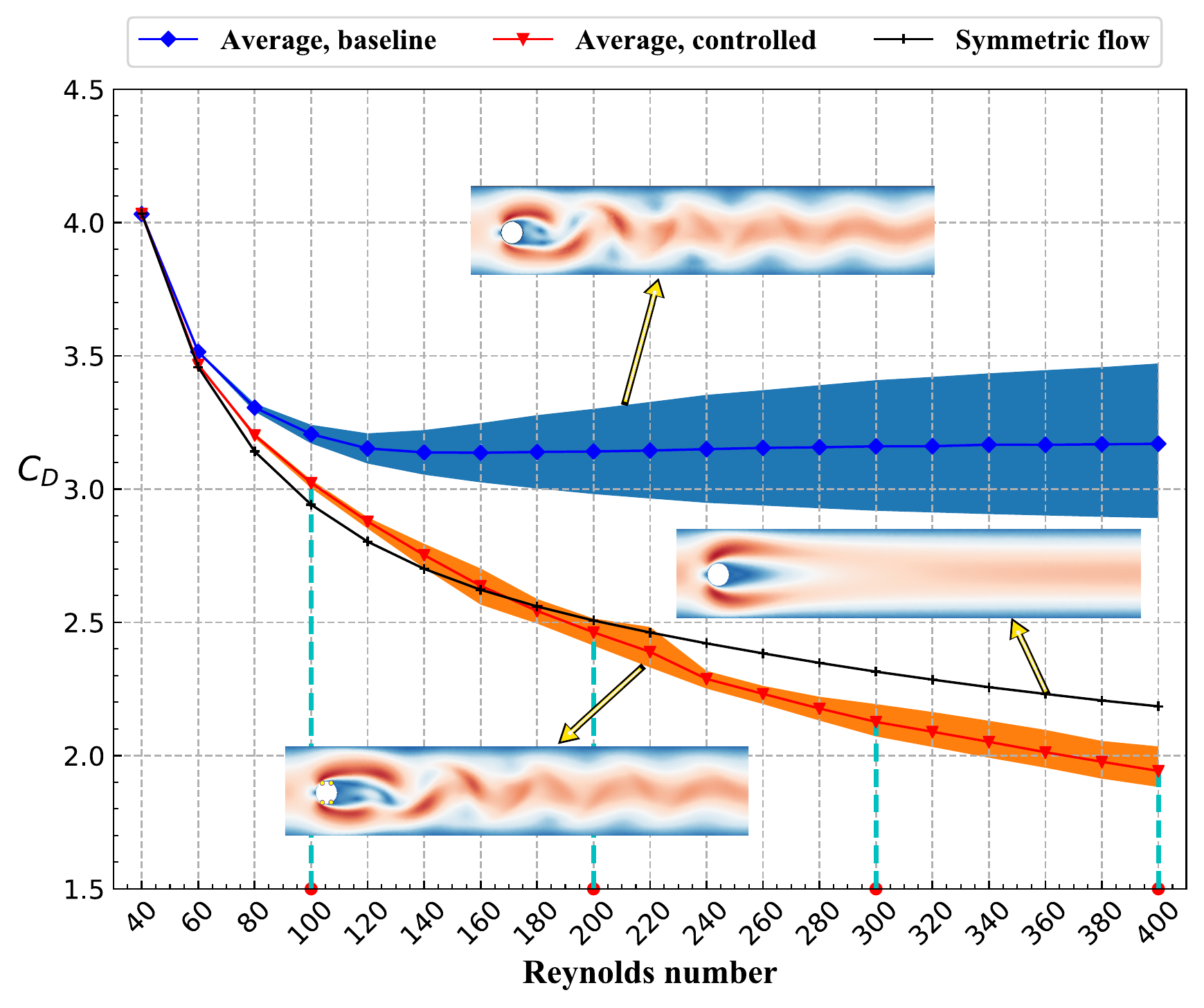}
  \caption{Average of the drag coefficient for flow with and without control at
  different Reynolds numbers, and corresponding drag coefficient using a
  symmetric boundary condition at the equatorial plane of flow domain. The
  shaded areas indicate the range of oscillation in each respective case when
  the flow appears to be pseudo-periodic with active flow control. The general
  active flow control strategy is discovered through training at 4 Reynolds
  number (highlighted by red dots): 100, 200, 300 and 400. The insets with
  velocity magnitude as contour represent the structure of corresponding flow
  state at a given $Re$. It is remarkable that control can be successfully
  applied at any $Re$ between 60 and 400. In addition, the values of the drag
  coefficient $C_D$ obtained with control compared with the symmetric case
  suggest that the global control strategy is close to being optimal on the
  range of Reynolds numbers considered \cite{Bergmann2005}.}
  \label{general_base_comparison}
\end{figure*}

In order to evaluate the efficiency of the control strategy obtained by the PPO
agent, the average values of the drag coefficient with active control are further
compared with the drag coefficient values obtained in the case where there is no
vortex shedding. Such flow state still exists in supercritical regime but it is
too unstable to be observed in experiments \cite{Protas2002}. However, it is easy to
obtain in numerical simulation, by using a symmetric boundary condition at the
equatorial plane of flow domain, similarly to what is performed in \cite{Rabault2019a}. As can be seen in Fig.\
\ref{general_base_comparison}, with the Reynolds number increasing, the drag
obtained at steady-state decreases (Symmetric flow curve). Relatively, the
contribution from the unsteady part to the drag becomes increasingly
significant. It is promising to see that the drag with active flow control is
even smaller than the drag obtained without vortex shedding if $Re\geq 200$,
indicating that the control strategy is close to the theoretical optimum
\cite{Bergmann2005}.

It is worth emphasizing that only four values of the Reynolds numbers, i.e.,$Re = 100$,
200, 300 and 400 (highlighted by red dots in Fig.\
\ref{general_base_comparison}) were used during the training process, while the
control is successful for any Re within that range (all markers on the figure correspond
to individual simulations where the PPO agent trained on only the 4 reference
Re values was used). In addition, the strategy is
still effective for active control even beyond the scope of $Re$ used for
training, for instance at $Re=80$. This, again, highlights the generalization
ability of the ANN and is of great importance for practical applications.

\section{Conclusions}\label{sec:conclusions}

In this study, the framework initially presented in the work of Rabault
\textit{et al.} \cite{Rabault2019a, Rabault2019b} is extended by demonstrating
the robustness and generalization ability of the PPO algorithm for
machine-learning-based AFC. This state-of-the-art DRL method can allow ANNs to
discover global active control strategies for flows over a range of Reynolds
numbers. An alternative smoothing law performing linear interpolation between
two successive actions is proposed to make the control values, i.e., the mass
flow rates of the synthetic jets, change smoothly in time. With this method, the
lift coefficient is made continuous to avoid non-physical jumps potentially
occurring at action updates. The learning environment used for training supports
four flow configurations with Reynolds numbers 100, 200, 300 and 400,
respectively. After training, the ANN is able to actively control the flow and
to reduce the drag by approximately 5.7\%, 21.6\%, 32.7\%, and 38.7\%, when
$Re$=100, 200, 300, and 400 respectively. More importantly, the ANN can also
effectively reduce drag for any previously unseen Re in the range from 60 to
400. By observing the flow field through its mean pressure and vorticity, one can
observe that the size of the separated wake and the vortex shedding area behind
the cylinder are enlarged, resulting in a reduction of the pressure drop behind
the cylinder and the oscillation frequency induced by the vortices. It should be
emphasized that only four values of $Re$ were used during the training process,
while the control is successful for any $Re$ in the range 60-400, which
highlights the generalization ability of the ANN and is of great importance for
practical applications. The averaged drag with control is further compared with
the drag value when using a symmetric boundary condition at the equatorial plane
of the flow domain. It is promising that the drag of the controlled flow is even
smaller than this symmetric baseline value if $Re\geq 200$, suggesting that the
control strategy is close to the theoretical optimum \cite{Bergmann2005}.
Moreover, the results indicate that in order to obtain better control
performance for more complex systems, like the flow at higher $Re$ in the present
case, embedding within the environment a number of systems with relatively
simple but similar properties seems to be an efficient strategy. This is similar to the
idea of transfer learning.

It should be noted that due to exploration noise and randomness involved in the
training process, the strategy discovered through ANNs together with the PPO
method may perform slightly different control performance in different training runs. However,
the qualitative strategies found are relatively similar from one run to another.

Despite the relative simplicity of the selected problem, the experience and
insights gained from this work are of great importance for progressing toward
the application of DRL to more practical engineering problems in fluid
mechanics. Although the computational cost remains a challenge to wide application
of DRL within Fluid Mechanics,
this challenge can be progressively solved owing to the rapid advancement of high-performance
computing architectures. Therefore, it is anticipated that significantly more
complex problems, such as instabilities in boundary
layers \cite{xu_mughal_gowree_atkin_sherwin_2017, xu_lombard_sherwin_2017}, can be tackled using methodologies based on the present work, possibly
in combination with other results and technical improvements such as the
encoding of physical invariance of the system to control within the ANN
architecture \cite{doi:10.1063/1.5132378, anderson2019cormorant}, or the identification of reduced-order, hidden features of
these systems \cite{doi:10.1063/5.0002051, Raissi1026}.
In order to support the further development of DRL applications in the fluid
mechanics community, all code used is released as
open-source (see Appendix \ref{appendix_a}).

\begin{acknowledgements}
This work is supported by the National Key Research and Development Program
(No.2019YFB1503701-02), the Funding of Nanjing Institute of Technology
(No.YKJ201943) and the Priority Academic Program Development of Jiangsu Higher
Education Institutions. Y. Wang acknowledges the support of the Natural Science
Foundation of China (Grant No. 11902153), National Numerical Wind Tunnel Project
(Grant No. NNW2019ZT2-B28), the Natural Science Foundation of Jiangsu Province
(Grant No. BK2019043306).
\end{acknowledgements}

\section{Data Availability Statement}
The data that support the findings of this study will be openly available at
 \url{https://github.com/thw1021/Cylinder2DFlowControlGeneral} upon publication
 in the peer-reviewed journal.

\appendix
\section{Open source code}\label{appendix_a}

The source code of this projcet together with all needed packages will released
at \url{https://github.com/thw1021/Cylinder2DFlowControlGeneral} upon
publication in the peer-reviewed journal. The CFD solver is built on the
open-source finite element package FEniCS \cite{Logg2012}. The DRL agent is
based on the open-source framework Tensorforce \cite{tensorforce}. The present
work is based on the multi-environment approach proposed by Rabault and Kuhnle
\cite{Rabault2019b} and the reader can also refer to the open source code
\url{https://github.com/jerabaul29/Cylinder2DFlowControlDRLParallel}.

\section{Evaluation of momentum injected into the flow field using 4 jets}\label{appendix_b}

When using 4 jets as schematically presented in Fig. \ref{BC}, some extra
momentum may be injected into the flow field. In this appendix, a mathematical
formulation is derived to evaluate the injected momentum.

The momentum injected into the flow field per unit time by the $i-th$ ($i=1,
2, 3, 4$) jet on horizontal direction can be evaluated as following:

\begin{eqnarray}
M_x^i &=& \int_{\theta_{0}^i-\omega/2}^{\theta_{0}^i+\omega/2}\rho u_{jet}(\theta;Q_i)u_{jet}(\theta;Q_i)\cos\theta\cdot \frac{D}{2}d\theta \nonumber \\
&=&\frac{\pi^4}{\omega^2(4\pi^2-\omega^2)}\cdot\frac{2\rho sin\frac{\omega}{2}}{D}\cdot Q_i^2\cos\theta_0^i.
\label{x_momentum}
\end{eqnarray}
where $\theta_0^i$ is the position of $i-th$ ($i=1, 2, 3, 4$) jet.

Consequently, the total momentum injected by 4 jets on horizontal direction is

\begin{eqnarray}
M_x&=&\sum_{i=1}^{4}M_x^i \nonumber \\
&=&\frac{2\rho\pi^4 sin\frac{\omega}{2}}{\omega^2(4\pi^2-\omega^2)D}\sum_{i=1}^{4}Q_i^2\cos\theta_0^i \nonumber \\
&=&\frac{2\rho\pi^4 sin\frac{\omega}{2}}{\omega^2(4\pi^2-\omega^2)D}(Q_1^2-Q_2^2-Q_3^2+Q_4^2)\cos75^\circ.
\label{total_momentum}
\end{eqnarray}

In the following, the injected momentum on horizontal direction is normalized
as follows:

\begin{equation}
M^*=M_x/M_{ref},
\end{equation}

\noindent where $M_{ref}=\int_{-D/2}^{D/2}\rho u_{inlet}(y)u_{inlet}(y)dy$ is
the reference momentum intercepting the cylinder.

\begin{figure}
\centering
\includegraphics[width=7.5cm, height=6.5cm]{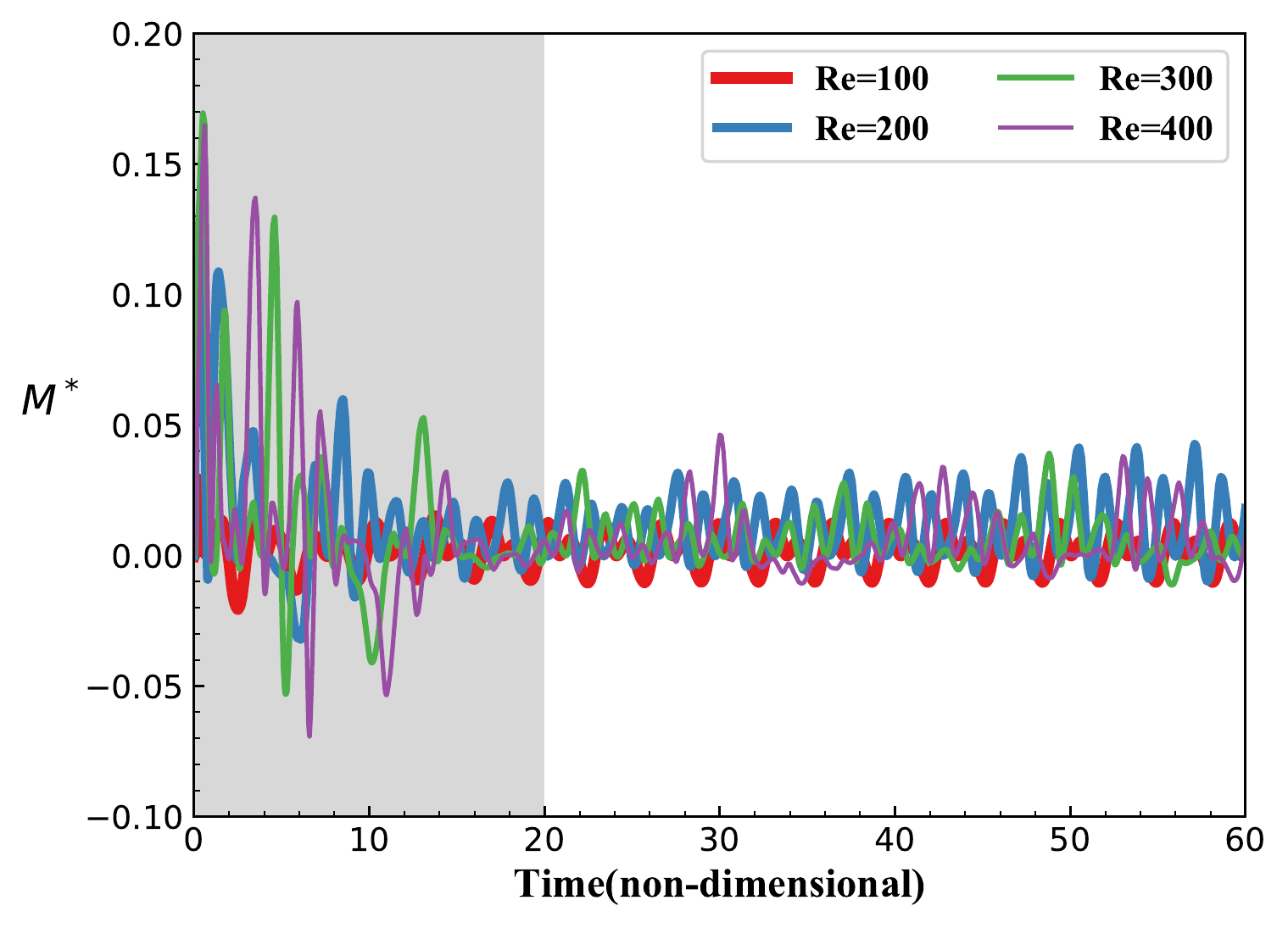}
\caption{Time series of normalized momentum horizontally injected into the flow
field by the 4 control jets at $Re=100$, 200, 300, 400.}
\label{fig:total_momentum}
\end{figure}

The time-resolved value of the normalized momentum added by the 4 jets when
applying the active flow control strategy to typical flow environments, is shown
in Fig.\ \ref{fig:total_momentum}. Obviously, more momentum will be injected for
flow at higher Reynolds number.

\bibliographystyle{unsrt}
\bibliography{aipsamp}% Produces the bibliography via BibTeX.

\begin{thebibliography}{10}

\bibitem{gad2000flow}
Mohamed Gad-el Hak.
\newblock {\em Flow Control: Passive, Active, and Reactive Flow Management}.
\newblock Cambridge University Press, 2000.

\bibitem{prandtl1904}
Ludwig Prandtl.
\newblock {\"U}ber fl{\"u}ssigkeitsbewegung bei sehr kleiner reibung.
\newblock In {\em Proceedings of the 3rd International Mathematics Congress},
  pages 484--491, Heidelberg, Germany, 1904.

\bibitem{articleRVinu}
Ricardo Vinuesa, Hossein Azizpour, Iolanda Leite, Madeline Balaam, Virginia
  Dignum, Sami Domisch, Anna Felländer, Simone Langhans, Max Tegmark, and
  Francesco Nerini.
\newblock The role of artificial intelligence in achieving the sustainable
  development goals.
\newblock {\em Nature Communications}, 11:233, 01 2020.

\bibitem{Shahrabi2019}
A.~F. Shahrabi.
\newblock The control of flow separation: Study of optimal open loop
  parameters.
\newblock {\em Physics of Fluids}, 31(3):035104, 2019.

\bibitem{doi:10.2514/1.J056258}
D.~Dolgopyat and A.~Seifert.
\newblock Active flow control virtual maneuvering system applied to
  conventional airfoil.
\newblock {\em AIAA Journal}, 57(1):72--89, 2019.

\bibitem{Zhu2019}
Hongjun Zhu, Tao Tang, Honglei Zhao, and Yue Gao.
\newblock Control of vortex-induced vibration of a circular cylinder using a
  pair of air jets at low reynolds number.
\newblock {\em Physics of Fluids}, 31(4):043603, 2019.

\bibitem{Wang2017}
Chenglei Wang, Hui Tang, Simon C.~M. Yu, and Fei Duan.
\newblock Control of vortex-induced vibration using a pair of synthetic jets:
  Influence of active lock-on.
\newblock {\em Physics of Fluids}, 29(8):083602, 2017.

\bibitem{doi:10.1063/1.5109320}
Baptiste Plumejeau, Sébastien Delprat, Laurent Keirsbulck, Marc Lippert, and
  Wafik Abassi.
\newblock Ultra-local model-based control of the square-back ahmed body wake
  flow.
\newblock {\em Physics of Fluids}, 31(8):085103, 2019.

\bibitem{WANG2016160}
Chenglei Wang, Hui Tang, Fei Duan, and Simon~C.M. Yu.
\newblock Control of wakes and vortex-induced vibrations of a single circular
  cylinder using synthetic jets.
\newblock {\em Journal of Fluids and Structures}, 60:160 -- 179, 2016.

\bibitem{doi:10.1063/1.5127202}
C.~Raibaudo, P.~Zhong, B.~R. Noack, and R.~J. Martinuzzi.
\newblock Machine learning strategies applied to the control of a fluidic
  pinball.
\newblock {\em Physics of Fluids}, 32(1):015108, 2020.

\bibitem{Aubrun2017}
Sandrine Aubrun, Annie Leroy, and Philippe Devinant.
\newblock {A review of wind turbine-oriented active flow control strategies}.
\newblock {\em Experiments in Fluids}, 58(10):134, 2017.

\bibitem{doi:10.1002/we.2109}
R.~Pereira, W.A. Timmer, G.~de~Oliveira, and G.J.W. van Bussel.
\newblock Design of hawt airfoils tailored for active flow control.
\newblock {\em Wind Energy}, 20(9):1569--1583, 2017.

\bibitem{doi:10.1002/we.1737}
Alexander Wolf, Thorsten Lutz, Werner Würz, Ewald Krämer, Oksana Stalnov, and
  Avraham Seifert.
\newblock Trailing edge noise reduction of wind turbine blades by active flow
  control.
\newblock {\em Wind Energy}, 18(5):909--923, 2015.

\bibitem{doi:10.2514/1.J056697}
Jeffrey Bons, Stuart Benton, Chiara Bernardini, and Matthew Bloxham.
\newblock Active flow control for low-pressure turbines.
\newblock {\em AIAA Journal}, 56(7):2687--2698, 2018.

\bibitem{Brunton2020}
Steven~L. Brunton, Bernd~R. Noack, and Petros Koumoutsakos.
\newblock Machine learning for fluid mechanics.
\newblock {\em Annual Review of Fluid Mechanics}, 52(1):477--508, 2020.

\bibitem{collis2004issues}
Scott Collis, Ronald~D. Joslin, Avi Seifert, and Vassilis Theofilis.
\newblock Issues in active flow control: theory, control, simulation, and
  experiment.
\newblock {\em Progress in Aerospace Sciences}, 40(4):237 -- 289, 2004.

\bibitem{doi:10.1063/1.4928896}
Thibault L.~B. Flinois and Tim Colonius.
\newblock Optimal control of circular cylinder wakes using long control
  horizons.
\newblock {\em Physics of Fluids}, 27(8):087105, 2015.

\bibitem{leclercq_2019}
Colin Leclercq, Fabrice Demourant, Charles Poussot-Vassal, and Denis Sipp.
\newblock Linear iterative method for closed-loop control of quasiperiodic
  flows.
\newblock {\em Journal of Fluid Mechanics}, 868:26–65, 2019.

\bibitem{Bergmann2005}
Michel Bergmann, Laurent Cordier, and Jean-Pierre Brancher.
\newblock Optimal rotary control of the cylinder wake using proper orthogonal
  decomposition reduced-order model.
\newblock {\em Physics of Fluids}, 17(9):097101, 2005.

\bibitem{cruz_wynn_rigas_morrison_2016}
R.~D. Brackston, J.~M. García de~la Cruz, A.~Wynn, G.~Rigas, and J.~F.
  Morrison.
\newblock Stochastic modelling and feedback control of bistability in a
  turbulent bluff body wake.
\newblock {\em Journal of Fluid Mechanics}, 802:726–749, 2016.

\bibitem{brunton2015closed}
Steven~L Brunton and Bernd~R Noack.
\newblock Closed-loop turbulence control: progress and challenges.
\newblock {\em Applied Mechanics Reviews}, 67(5):050801, 2015.

\bibitem{doi:10.1063/1.869789}
Wade Schoppa and Fazle Hussain.
\newblock A large-scale control strategy for drag reduction in turbulent
  boundary layers.
\newblock {\em Physics of Fluids}, 10(5):1049--1051, 1998.

\bibitem{gautier_aider_duriez_noack_segond_abel_2015}
N.~Gautier, J.-L. Aider, T.~Duriez, B. R. Noack, M.~Segond, and M.~Abel.
\newblock Closed-loop separation control using machine learning.
\newblock {\em Journal of Fluid Mechanics}, 770:442–457, 2015.

\bibitem{duriez2017machine}
Thomas Duriez, Steven~L Brunton, and Bernd~R Noack.
\newblock {\em Machine Learning Control-Taming Nonlinear Dynamics and
  Turbulence}.
\newblock Springer, 2017.

\bibitem{doi:10.1063/1.5115258}
Feng Ren, Chenglei Wang, and Hui Tang.
\newblock Active control of vortex-induced vibration of a circular cylinder
  using machine learning.
\newblock {\em Physics of Fluids}, 31(9):093601, 2019.

\bibitem{debien2016closed}
Antoine Debien, Kai~AFF Von~Krbek, Nicolas Mazellier, Thomas Duriez, Laurent
  Cordier, Bernd~R Noack, Markus~W Abel, and Azeddine Kourta.
\newblock Closed-loop separation control over a sharp edge ramp using genetic
  programming.
\newblock {\em Experiments in fluids}, 57(3):40, 2016.

\bibitem{Mnih2015}
Volodymyr Mnih, Koray Kavukcuoglu, David Silver, Andrei~A Rusu, Joel Veness,
  Marc~G Bellemare, Alex Graves, Martin Riedmiller, Andreas~K Fidjeland, Georg
  Ostrovski, Stig Petersen, Charles Beattie, Amir Sadik, Ioannis Antonoglou,
  Helen King, Dharshan Kumaran, Daan Wierstra, Shane Legg, and Demis Hassabis.
\newblock {Human-level control through deep reinforcement learning.}
\newblock {\em Nature}, 518(7540):529--33, 2015.

\bibitem{duan2016benchmarking}
Yan Duan, Xi~Chen, Rein Houthooft, John Schulman, and Pieter Abbeel.
\newblock Benchmarking deep reinforcement learning for continuous control.
\newblock In {\em International Conference on Machine Learning}, pages
  1329--1338, 2016.

\bibitem{gu2016continuous}
Shixiang Gu, Timothy Lillicrap, Ilya Sutskever, and Sergey Levine.
\newblock Continuous deep q-learning with model-based acceleration.
\newblock In {\em International Conference on Machine Learning}, pages
  2829--2838, 2016.

\bibitem{hessel2018rainbow}
Matteo Hessel, Joseph Modayil, Hado Van~Hasselt, Tom Schaul, Georg Ostrovski,
  Will Dabney, Dan Horgan, Bilal Piot, Mohammad Azar, and David Silver.
\newblock Rainbow: Combining improvements in deep reinforcement learning.
\newblock In {\em Thirty-Second AAAI Conference on Artificial Intelligence},
  2018.

\bibitem{mnih2013playing}
Mnih Volodymyr, Kavukcuoglu Koray, Silver David, Graves Alex, Antonoglou
  Ioannis, W~Daan, and R~Martin.
\newblock Playing atari with deep reinforcement learning.
\newblock In {\em NIPS Deep Learning Workshop}, 2013.

\bibitem{li2016deep}
Jiwei Li, Will Monroe, Alan Ritter, Dan Jurafsky, Michel Galley, and Jianfeng
  Gao.
\newblock Deep reinforcement learning for dialogue generation.
\newblock In {\em Proceedings of the 2016 Conference on Empirical Methods in
  Natural Language Processing}, pages 1192--1202, 2016.

\bibitem{gu2017deep}
Shixiang Gu, Ethan Holly, Timothy Lillicrap, and Sergey Levine.
\newblock Deep reinforcement learning for robotic manipulation with
  asynchronous off-policy updates.
\newblock In {\em 2017 IEEE International Conference on Robotics and Automation
  (ICRA)}, pages 3389--3396. IEEE, 2017.

\bibitem{Rabault_2017}
Jean Rabault, Jostein Kolaas, and Atle Jensen.
\newblock Performing particle image velocimetry using artificial neural
  networks: a proof-of-concept.
\newblock {\em Measurement Science and Technology}, 28(12):125301, Nov 2017.

\bibitem{MENDEZ2018256}
M.A. Mendez, M.T. Scelzo, and J.-M. Buchlin.
\newblock Multiscale modal analysis of an oscillating impinging gas jet.
\newblock {\em Experimental Thermal and Fluid Science}, 91:256 -- 276, 2018.

\bibitem{MENDEZ201948}
M.A. Mendez, A.~Gosset, and J.-M. Buchlin.
\newblock Experimental analysis of the stability of the jet wiping process,
  part {II}: Multiscale modal analysis of the gas jet-liquid film interaction.
\newblock {\em Experimental Thermal and Fluid Science}, 106:48 -- 67, 2019.

\bibitem{mendez_balabane_buchlin_2019}
M.~A. Mendez, M.~Balabane, and J.-M. Buchlin.
\newblock Multi-scale proper orthogonal decomposition of complex fluid flows.
\newblock {\em Journal of Fluid Mechanics}, 870:988–1036, 2019.

\bibitem{doi:10.1063/1.5144861}
Massoud Tatar and Mohammad~Hossein Sabour.
\newblock Reduced-order modeling of dynamic stall using neuro-fuzzy inference
  system and orthogonal functions.
\newblock {\em Physics of Fluids}, 32(4):045101, 2020.

\bibitem{guastoni2019prediction}
P.~A. Srinivasan, L.~Guastoni, H.~Azizpour, P.~Schlatter, and R.~Vinuesa.
\newblock Predictions of turbulent shear flows using deep neural networks.
\newblock {\em Physical Review Fluids}, 4:054603, May 2019.

\bibitem{doi:10.1063/1.5094943}
Vinothkumar Sekar, Qinghua Jiang, Chang Shu, and Boo~Cheong Khoo.
\newblock Fast flow field prediction over airfoils using deep learning
  approach.
\newblock {\em Physics of Fluids}, 31(5):057103, 2019.

\bibitem{doi:10.1063/1.5109698}
S.~Z. Islami~rad, R.~Gholipour~Peyvandi, and S.~Sadrzadeh.
\newblock Determination of the volume fraction in (water-gasoil-air) multiphase
  flows using a simple and low-cost technique: Artificial neural networks.
\newblock {\em Physics of Fluids}, 31(9):093301, 2019.

\bibitem{PhysRevFluids.4.093902}
Guido Novati, L.~Mahadevan, and Petros Koumoutsakos.
\newblock Controlled gliding and perching through deep-reinforcement-learning.
\newblock {\em Physical Review Fluids}, 4:093902, Sep 2019.

\bibitem{doi:10.1137/130943078}
Mattia. Gazzola, Babak. Hejazialhosseini, and Petros. Koumoutsakos.
\newblock Reinforcement learning and wavelet adapted vortex methods for
  simulations of self-propelled swimmers.
\newblock {\em SIAM Journal on Scientific Computing}, 36(3):B622--B639, 2014.

\bibitem{brauer_koumoutsakos_2016}
M.~Gazzola, A. A. Tchieu, D.~Alexeev, A.~de~Brauer, and P.~Koumoutsakos.
\newblock Learning to school in the presence of hydrodynamic interactions.
\newblock {\em Journal of Fluid Mechanics}, 789:726–749, 2016.

\bibitem{Verma5849}
Siddhartha Verma, Guido Novati, and Petros Koumoutsakos.
\newblock Efficient collective swimming by harnessing vortices through deep
  reinforcement learning.
\newblock {\em Proceedings of the National Academy of Sciences},
  115(23):5849--5854, 2018.

\bibitem{reddy2018glider}
Gautam Reddy, Jerome Wong-Ng, Antonio Celani, Terrence~J Sejnowski, and Massimo
  Vergassola.
\newblock Glider soaring via reinforcement learning in the field.
\newblock {\em Nature}, 562(7726):236, 2018.

\bibitem{PhysRevLett.118.158004}
Simona Colabrese, Kristian Gustavsson, Antonio Celani, and Luca Biferale.
\newblock Flow navigation by smart microswimmers via reinforcement learning.
\newblock {\em Physical Review Letters}, 118:158004, Apr 2017.

\bibitem{Rabault2019a}
Jean Rabault, Miroslav Kuchta, Atle Jensen, Ulysse R{\'{e}}glade, and Nicolas
  Cerardi.
\newblock {Artificial neural networks trained through deep reinforcement
  learning discover control strategies for active flow control}.
\newblock {\em Journal of Fluid Mechanics}, 865:281--302, 2019.

\bibitem{Rabault2019b}
Jean Rabault and Alexander Kuhnle.
\newblock {Accelerating deep reinforcement learning strategies of flow control
  through a multi-environment approach}.
\newblock {\em Physics of Fluids}, 31(9):094105, 2019.

\bibitem{Schafer1996}
M~Sch{\"{a}}fer, S~Turek, F~Durst, E~Krause, and R~Rannacher.
\newblock {Benchmark Computations of Laminar Flow Around a Cylinder BT - Flow
  Simulation with High-Performance Computers II: DFG Priority Research
  Programme Results 1993–1995}.
\newblock pages 547--566. Vieweg Teubner Verlag, Wiesbaden, 1996.

\bibitem{Goda1979}
Katuhiko Goda.
\newblock {A multistep technique with implicit difference schemes for
  calculating two- or three-dimensional cavity flows}.
\newblock {\em Journal of Computational Physics}, 30(1):76--95, 1979.

\bibitem{Logg2012}
Anders Logg, Kent-Andre Mardal, and Garth Wells.
\newblock {\em {Automated solution of differential equations by the finite
  element method: The FEniCS book}}, volume~84.
\newblock Springer Science {\&} Business Media, 2012.

\bibitem{Davis:1997:UMM:258211.258225}
Timothy~A. Davis and Iain~S. Duff.
\newblock An unsymmetric-pattern multifrontal method for sparse lu
  factorization.
\newblock {\em SIAM Journal on Matrix Analysis and Applications},
  18(1):140--158, 1997.

\bibitem{Kutz2017}
J.~Nathan Kutz.
\newblock {Deep learning in fluid dynamics}.
\newblock {\em Journal of Fluid Mechanics}, 814:1--4, 2017.

\bibitem{Thuerey2019}
Nils Thuerey, Konstantin Wei{\ss}enow, Lukas Prantl, and Xiangyu Hu.
\newblock {Deep Learning Methods for Reynolds-Averaged Navier–Stokes
  Simulations of Airfoil Flows}.
\newblock {\em AIAA Journal}, pages 1--12, Nov 2019.

\bibitem{Beck2019}
Andrea Beck, David Flad, and Claus-Dieter Munz.
\newblock {Deep neural networks for data-driven LES closure models}.
\newblock {\em Journal of Computational Physics}, 398:108910, 2019.

\bibitem{Sirignano2018}
Justin Sirignano and Konstantinos Spiliopoulos.
\newblock {DGM: A deep learning algorithm for solving partial differential
  equations}.
\newblock {\em Journal of Computational Physics}, 2018.

\bibitem{Raissi2019}
M~Raissi, P~Perdikaris, and G~E Karniadakis.
\newblock {Physics-informed neural networks: A deep learning framework for
  solving forward and inverse problems involving nonlinear partial differential
  equations}.
\newblock {\em Journal of Computational Physics}, 378:686--707, 2019.

\bibitem{Yan2019}
Xinghui Yan, Jihong Zhu, Minchi Kuang, and Xiangyang Wang.
\newblock {Aerodynamic shape optimization using a novel optimizer based on
  machine learning techniques}.
\newblock {\em Aerospace Science and Technology}, 86:826--835, 2019.

\bibitem{Yonekura2019}
Kazuo Yonekura and Hitoshi Hattori.
\newblock {Framework for design optimization using deep reinforcement
  learning}.
\newblock {\em Structural and Multidisciplinary Optimization}, 60:1709--1713,
  2019.

\bibitem{Silver2016}
David Silver, Aja Huang, Chris~J Maddison, Arthur Guez, Laurent Sifre, George
  van~den Driessche, Julian Schrittwieser, Ioannis Antonoglou, Veda
  Panneershelvam, Marc Lanctot, Sander Dieleman, Dominik Grewe, John Nham, Nal
  Kalchbrenner, Ilya Sutskever, Timothy Lillicrap, Madeleine Leach, Koray
  Kavukcuoglu, Thore Graepel, and Demis Hassabis.
\newblock {Mastering the game of Go with deep neural networks and tree search}.
\newblock {\em Nature}, 529(7587):484--489, 2016.

\bibitem{8798254}
E.~{Bøhn}, E.~M. {Coates}, S.~{Moe}, and T.~A. {Johansen}.
\newblock Deep reinforcement learning attitude control of fixed-wing uavs using
  proximal policy optimization.
\newblock In {\em 2019 International Conference on Unmanned Aircraft Systems
  (ICUAS)}, pages 523--533, June 2019.

\bibitem{Sutton2018}
Richard~S Sutton and Andrew~G Barto.
\newblock {\em {Reinforcement learning: An introduction}}.
\newblock MIT Press, 2018.

\bibitem{Recht2019}
Benjamin Recht.
\newblock A tour of reinforcement learning: The view from continuous control.
\newblock {\em Annual Review of Control, Robotics, and Autonomous Systems},
  2(1):253--279, 2019.

\bibitem{braylan:aaai15ws}
Alexander Braylan, Mark Hollenbeck, Elliot Meyerson, and Risto Miikkulainen.
\newblock Frame skip is a powerful parameter for learning to play atari.
\newblock In {\em AAAI-15 Workshop on Learning for General Competency in Video
  Games}, 2015.

\bibitem{alex2018adaptive}
Alexander Neitz, Giambattista Parascandolo, Stefan Bauer, and Bernhard
  Sch{\"o}lkopf.
\newblock Adaptive skip intervals: Temporal abstraction for recurrent dynamical
  models.
\newblock In {\em Advances in Neural Information Processing Systems}, pages
  9816--9826, 2018.

\bibitem{Protas2002}
B.~Protas and J.~E. Wesfreid.
\newblock Drag force in the open-loop control of the cylinder wake in the
  laminar regime.
\newblock {\em Physics of Fluids}, 14(2):810--826, 2002.

\bibitem{taylor2009transfer}
Matthew~E Taylor and Peter Stone.
\newblock Transfer learning for reinforcement learning domains: A survey.
\newblock {\em Journal of Machine Learning Research}, 10(Jul):1633--1685, 2009.

\bibitem{xu_mughal_gowree_atkin_sherwin_2017}
Hui Xu, Shahid~M. Mughal, Erwin~R. Gowree, Chris~J. Atkin, and Spencer~J.
  Sherwin.
\newblock Destabilisation and modification of tollmien–schlichting
  disturbances by a three-dimensional surface indentation.
\newblock {\em Journal of Fluid Mechanics}, 819:592–620, 2017.

\bibitem{xu_lombard_sherwin_2017}
Hui Xu, Jean-Eloi~W. Lombard, and Spencer~J. Sherwin.
\newblock Influence of localised smooth steps on the instability of a boundary
  layer.
\newblock {\em Journal of Fluid Mechanics}, 817:138–170, 2017.

\bibitem{doi:10.1063/1.5132378}
Vincent Belus, Jean Rabault, Jonathan Viquerat, Zhizhao Che, Elie Hachem, and
  Ulysse Reglade.
\newblock Exploiting locality and translational invariance to design effective
  deep reinforcement learning control of the 1-dimensional unstable falling
  liquid film.
\newblock {\em AIP Advances}, 9(12):125014, 2019.

\bibitem{anderson2019cormorant}
Brandon Anderson, Truong~Son Hy, and Risi Kondor.
\newblock Cormorant: Covariant molecular neural networks.
\newblock In {\em Advances in Neural Information Processing Systems}, pages
  14510--14519, 2019.

\bibitem{doi:10.1063/5.0002051}
Suraj Pawar, Shady~E. Ahmed, Omer San, and Adil Rasheed.
\newblock Data-driven recovery of hidden physics in reduced order modeling of
  fluid flows.
\newblock {\em Physics of Fluids}, 32(3):036602, 2020.

\bibitem{Raissi1026}
Maziar Raissi, Alireza Yazdani, and George~Em Karniadakis.
\newblock Hidden fluid mechanics: Learning velocity and pressure fields from
  flow visualizations.
\newblock {\em Science}, 367(6481):1026--1030, 2020.

\bibitem{tensorforce}
Alexander Kuhnle, Michael Schaarschmidt, and Kai Fricke.
\newblock Tensorforce: a tensorflow library for applied reinforcement learning.
\newblock https://github.com/tensorforce/tensorforce.

\end{thebibliography}

\end{document}